\documentclass[]{emulateapj}

\usepackage{CJK}
\usepackage{amsmath}
\usepackage{natbib}
\usepackage{mathrsfs}
\usepackage{epstopdf}
\usepackage{txfonts}
\usepackage{graphicx,color}

\newcommand{\cs}{c_{\rm s}}
\newcommand{\Dr}{\Delta r}
\newcommand{\m}[1]{{#1}_{m}}
\newcommand{\dr}{{d}r}
\newcommand{\eqnref}[1]{Equation \ref{#1}}
\newcommand{\secref}[1]{Section \ref{#1}}
\newcommand{\figref}[1]{Figure \ref{#1}}
\newcommand{\tabref}[1]{Table \ref{#1}}

\newcommand{\Bo}{\beta}
\newcommand{\ddr}[1]{\frac{{d}#1}{\dr}}
\newcommand{\ddlnr}[1]{\frac{{d}#1}{{\rm d}\ln{r}}}
\newcommand{\ddrr}[1]{\frac{{d^2} #1}{\dr^2}}
\newcommand{\ppr}[1]{\frac{\partial#1}{\partial r}}

\shorttitle{Irradiation Instability at the Inner Edges of Accretion Disks}
\shortauthors{Fung \& Artymowicz}

\begin{document}
\begin{CJK*}{UTF8}{bsmi}
\title{Irradiation Instability at the Inner Edges of Accretion Disks}

\author{Jeffrey Fung (馮澤之)\altaffilmark{1} and Pawel Artymowicz\altaffilmark{1,2}}

\altaffiltext{1}{Department of Astronomy and Astrophysics, University of Toronto, 50 St. George Street, Toronto, ON, Canada M5S 3H4}
\altaffiltext{2}{Department of Physical and Environmental Sciences, University of Toronto at Scarborough, 1265 Military Trail, Scarborough, ON, Canada M1C 1A4}

\email{fung@astro.utoronto.ca}

\begin{abstract}

An instability can potentially operate in highly irradiated disks where the disk sharply transitions from being radially transparent to opaque (the ''transition region''). Such conditions may exist at the inner edges of transitional disks around T Tauri stars and accretion disks around AGNs. We derive the criterion for this instability, which we term the ''irradiation instability'', or IRI. We also present the linear growth rate as a function of $\Bo$, the ratio between radiation force and gravity, and $\cs$, the sound speed of the disk, obtained using two methods: a semi-analytic analysis of the linearized equations and a numerical simulation using the GPU-accelerated hydrodynamical code \texttt{PEnGUIn}. In particular, we find that IRI occurs at $\Bo \sim 0.1$ if the transition region extends as wide as $\sim 0.05r$, and at higher $\Bo$ values if it is wider. This threshold value applies to $\cs$ ranging from $3\%$ of the Keplerian orbital speed to $5\%$, and becomes higher if $\cs$ is lower. Furthermore, in the nonlinear evolution of the instability, disks with a large $\Bo$ and small $\cs$ exhibit ''clumping'', extreme local surface density enhancements that can reach over ten times the initial disk surface density.

\end{abstract}
%------------------------------------------------------------

\keywords{accretion, accretion disks, hydrodynamics, instabilities, protoplanetary disks, radiation: dynamics}

%==================================================================
\section{Introduction}
\label{sec:intro}
%==================================================================

Accretion disks are susceptible to a wide range of instabilities, including 
the magnetorotational instability (MRI) \citep{MRI}, gravitational 
instability \citep{GI2,GI}, Papaloizou-Pringle instability \citep{PP84,PP85,PP87,GGN86}, 
and Rossby wave instability (RWI) \citep{RWI1}. The list goes on as non-ideal 
MHD and vertical shearing \citep{VShear} are considered. These 
instabilities drive the evolution of disks by generating turbulence 
and creating complex, sometimes extreme, structures, such as the formation 
of planets in protoplanetary disks.

Radiation pressure is a force generally present in all types of accretion 
disks. Its effect on accretion disks has been studied in many different aspects, 
including driving disk winds in active galactic nuclei (AGN) \citep[e.g.][]{Higginbottom2014},
shaping particle size distributions in debris disks \citep{Thebault2014}, and 
influencing the motions of the inner rims of transitional disks \citep{CM07,DD11}. 
We demonstrate in this paper that radiation pressure can also cause a disk instability of its own 
kind. In the following, we give a brief introduction to this instability before 
launching into the formal theoretical work.

The strength of radiation pressure compared to gravity is measured by the number $\Bo$:
\begin{equation}
\label{eq:beta0}
\Bo = \frac{\kappa_{\rm opa} L}{4 \pi cGM} ~,
\end{equation}
where $L$ is the central object's luminosity, $M$ is its mass, and 
$\kappa_{\rm opa}$ is the opacity of the disk material; $c$ and $G$ are the 
speed of light and gravitational constant respectively. The key to this instability is 
shadowing. As the front part of the disk gets pushed by radiation pressure, it also 
casts a shadow that reduces the amount of radiation pressure on the material further out in the disk. 
In a 1D, radial picture, since radiation pressure always diminishes outward, 
the inner part of a disk always feels a stronger push than the outer part, and the 
net effect is therefore radial compression. In other words, any two concentric disk 
annuli would feel an attraction between them due to the combined effects of radiation 
pressure and shadowing.

\begin{figure}[]
\includegraphics[width=0.99\columnwidth]{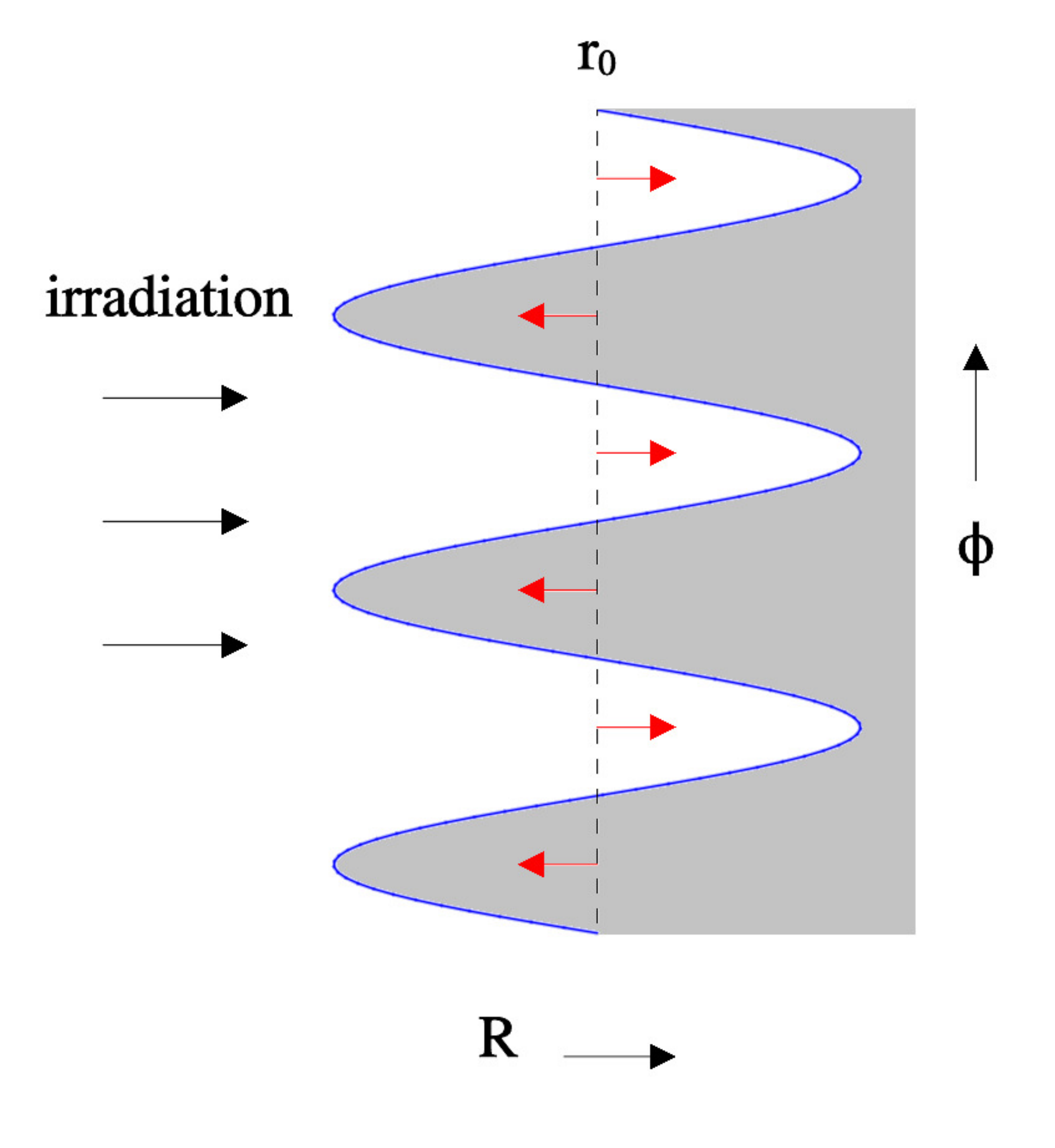}
\caption{Simple illustration describing IRI. The blue curve denotes the orbit of a perturbed disk element oscillating around its guiding center, denoted by the dashed black line at $r_0$. The shaded area is where the disk sees the shadow cast by the perturbed element. The red arrows show the directions of radial forcing on the background disk relative to the average amount of radiation pressure received along $r_0$. These arrows are inward when they are in the shadow of the element, and outward when they are not. One can see that the background disk near $r_0$ is forced in the direction of amplifying the initial perturbation.}
\label{fig:iri}
\end{figure}

This 1D scenario does not easily extend to a 2D disk however, because radiation 
pressure from a central source does not exert any azimuthal force. By the conservation 
of angular momentum, when a disk element is perturbed radially, it will oscillate at 
some epicyclic frequency. \figref{fig:iri} illustrates what effect this oscillating 
element has on the disk. One can see that disk material near the orbit of the perturbed 
element will experience a variation in shadowing along the azimuth. This variation creates 
a forcing that induces the unperturbed material to follow the motion of the perturbed element.
The result is a global collective motion that is capable of growing on its own. We term 
this phenomenon the ''irradiation instability'' (IRI), since it relies on irradiation 
by the central object.

Because a larger $\Bo$ allows for a more rapid radial motion, its value is crucial for the 
survival of this collective motion against disk shear. In most systems, dust grains provide 
the largest contribution to $\Bo$. In circumstellar disks, micron-size grains 
can have $\Bo>1$ for F-type stars, and up to $\Bo\sim10^1$ for A-type stars (e.g., 
Equation 10 of \citet{KW2013}). Given that the gas-to-dust ratio is typically 
$\sim10^2$, $\Bo$ of a perfectly coupled gas+dust mixture may be of the order of a few percent. 
Additionally, dust settling can enhance $\Bo$ in the midplane by reducing the local 
gas-to-dust ratio, while the radial migration of dust results in size segregation
 \citep{Thebault2014}, which can also enhance $\Bo$ at local radii. In other systems 
where radiation pressure can drive significant mass loss, such as AGN accretion disks, one 
would even expect $\Bo$ to exceed unity.

This paper aims to provide a basic understanding of IRI, of both the conditions 
that trigger it, and its consequences. In \secref{sec:theory}, we present a 
theoretical foundation for IRI and derive its instability criterion. \secref{sec:disk} 
contains our disk model. \secref{sec:two} describes our semi-analytic and
numerical methods. \secref{sec:result} reports the modal growth rate as a function of $\Bo$ and the 
sound speed $\cs$ of the disk, and gives a discussion on the nonlinear evolution of IRI. 
\secref{sec:disc} concludes with an outlook for future work.

%==================================================================
\section{The Linear Theory}
\label{sec:theory}
%==================================================================
We follow the method of \citet{GT1979}, using similar notation, to derive the linear response of a 2D disk stirred by radiation pressure. We start with the continuity equation and the conservation of momentum:

\begin{align}
\label{eq:cont}
\frac{\partial\Sigma}{\partial t} + {\bf\nabla}\cdot(\Sigma{\bf v}) &= 0,\\
\label{eq:moment1}
\frac{\partial{\bf v}}{\partial t} + ({\bf v}\cdot{\bf\nabla}){\bf v} &= -{\bf\nabla}\eta -\frac{GM\left(1-\Bo e^{-\tau}\right)}{r^2}\hat{\bf r},
\end{align}
where $\Sigma$ is the surface density of the disk; ${\bf v}$ is the 2D velocity field; $\eta$ is the specific enthalpy such that ${\bf\nabla}\eta = {\bf\nabla}P/\Sigma$, where $P$ is the vertically averaged gas pressure; and $\tau$ is the optical depth of the disk. We denote the Keplerian orbital frequency as $\Omega_{k}$, and the sound speed $\cs$ is defined by the ideal gas law $P=\cs^2 \Sigma$. $\tau$ depends on the density distribution by the following equation:

\begin{equation}
\tau = \int^r_0 \kappa_{\rm opa}\rho \dr' ~,
\end{equation}
where $\rho$ is the density of the disk. Near the midplane, $\rho\propto \Sigma/h$, where $h = \cs/\Omega_{k}$ is the scale height of the disk. Note that with \eqnref{eq:moment1} we have neglected the scattering of light into the azimuthal direction.

$\Sigma$, $\eta$, and $\bf v$ can be separated into a background quantity (without any subscript) and a perturbed quantity (denoted by the subscript ''$m$''). We assume the background disk to be axisymmetric and in hydrostatic equilibrium so that ${\bf v}=$($0,r\Omega$), where
\begin{equation}
\label{eq:omg}
\Omega = \sqrt{\Omega_{k}^2 \left(1-\Bo e^{-\tau}\right)+\frac{1}{r}\ddr{\eta}} ~,
\end{equation}
and the components of the perturbed velocity are denoted as $\m{\bf v}\equiv$($u,\upsilon$). To simplify the notation, we also define the background and perturbed radiation force as $F$ and $\m{F}$: 
\begin{align}
F &= r \Omega_{k}^2 \Bo e^{-\tau},\\
\m{F} &= -F\int^r_0 \frac{\m{\eta}}{\cs^2}\frac{{\rm d}\tau}{\dr'}\dr' ~.
\end{align}

For a small perturbation, it follows from Equations \ref{eq:cont} and \ref{eq:moment1} that the perturbed quantities are governed by the following linearized equations:
\begin{align}
\label{eq:cont2}
\frac{\partial\m{\Sigma}}{\partial t} + {\bf\nabla}(\Sigma\m{\bf v}) + {\bf\nabla}(\m{\Sigma}{\bf v}) = &~0 ~,\\
\label{eq:moment4}
\frac{\partial\m{\bf v}}{\partial t} + ({\bf v}\cdot{\bf\nabla})\m{\bf v} + (\m{\bf v}\cdot{\bf\nabla}){\bf v} = &-{\bf\nabla}\m{\eta} + \m{F} \hat{\bf r} ~.
\end{align}
Without a loss of generality, we can assume a form of the solution for the perturbed quantities $\m{\Sigma}$, $\m{\eta}$, $u$ and $\upsilon$:
\begin{equation}
\label{eq:soln}
\m{X}(r,\theta,t)=X(r)e^{i\left(m\theta-\omega t\right)} ~,
\end{equation}
for some complex function $X(r)$ and complex number $\omega$, while $m$ is the azimuthal mode number. Substituting this form into \eqnref{eq:moment4}, we find
\begin{align}
\label{eq:u}
u &= -\frac{i}{D}\left[\frac{2m\Omega}{r}\m{\eta} + \m{\Omega}\left(\ppr{\m{\eta}}-\m{F}\right)\right] ~,\\
\label{eq:nu}
\upsilon &= \frac{1}{D}\left[\frac{m\m{\Omega}}{r}\m{\eta} + 2\left(\Omega+\frac{r}{2}\ddr{\Omega}\right)\left(\ppr{\m{\eta}}-\m{F}\right)\right] ~.
\end{align}
The pattern rotation frequency $\m{\Omega}$ and the coefficient $D$ are defined as
\begin{align}
\label{eq:patfreq}
\m{\Omega}&\equiv m\Omega-\omega ~,\\
\label{eq:D}
D&\equiv\kappa^2-\m{\Omega}^2 ~,\\
\label{eq:kappa}
\kappa^2&= \frac{1}{r^3}\ddr{\left[r^4 \Omega^2\right]} ~,
\end{align}
where $\kappa$ is the epicyclic frequency of the unperturbed orbit. To solve for $\m{\Sigma}$, or equivalently $\m{\eta}$, we substitute \eqnref{eq:u} and \ref{eq:nu} into \eqnref{eq:cont2}, giving
\begin{equation}
\label{eq:2ndeq}
\frac{\partial^2\m{\eta}}{\partial r^2}+a(r)\ppr{\m{\eta}}+b(r)\m{\eta}+c(r)\int^r_0 \frac{\m{\eta}}{\cs^2}\frac{{\rm d}\tau}{\dr'} \dr' = 0,
\end{equation}
where
\begin{align}
\nonumber
a &\equiv \frac{\partial}{\partial r}\ln{\left(\frac{r\Sigma}{D}\right)} ~,\\
\nonumber
b &\equiv \frac{2m\Omega}{r \m{\Omega}}\ppr{}\ln{\left(\frac{\Sigma \Omega_0}{D}\right)}-\frac{m^2}{r^2}+\frac{1}{\cs^2}\left(F\ddr{\tau}-D\right) ~,\\
\nonumber
c &\equiv F \left(\ppr{}\ln{\left(\frac{r\Sigma F}{D}\right)}-\frac{2m\Omega}{r\m{\Omega}}\right) ~.
\end{align}
We arrive at a second-order integro-differential equation for $\m{\eta}$.

%==================================================================
\subsection{Instability Criterion}
\label{sec:crit}
%==================================================================
A local criterion for axisymmetric instability can be derived from \eqnref{eq:2ndeq}. We apply the WKB approximation and write $\m{\eta}\sim e^{i\int^r_0 k_{r} \dr'}$, where $k_{r}\gg \frac{1}{r}$ is the radial wave number. We then separate the real and imaginary part of the equation.
Finally, setting $m=0$, the dispersion relation can be written as:
\begin{equation}
\label{eq:disper}
\omega^2 = \kappa^2 + k_{r}^2 \cs^2 - \Omega_{k}^2 \Bo e^{-\tau} \left( \ddlnr{\tau} + \m{\tilde\tau}\ddlnr{\ln{\left[r\mathscr{R}\right]}} \right) ~,
\end{equation}
where
\begin{align}
\mathscr{R} & \equiv \frac{\Sigma \Omega_k \Bo e^{-\tau}}{\kappa^2} ~,\\
\label{eq:tau_m}
\m{\tilde\tau} & \equiv \m{\tau}\left(\frac{\m{\Sigma}}{\Sigma}\right)^{-1} = \frac{\cs^2}{\m{\eta}} \int^r_0 \frac{\m{\eta}}{\cs^2}\frac{{\rm d}\tau}{\dr'} \dr' ~.
\end{align}
$\mathscr{R}$ has the same units as the inverse of vortensity, but is a quantity that depends on radiation pressure. $\m{\tilde\tau}$ is the ratio between the perturbed optical depth $\m{\tau}$ and the relative surface density perturbation $\m{\Sigma}/\Sigma$. The local disk is unstable if a solution for $k_{r}$ exists given $\omega^2=0$, which denotes the line of neutral stability. Setting $\omega^2=0$, the condition for $k_{r}^2>0$ is
\begin{equation}
\label{eq:crit1}
\Bo e^{-\tau} \left(\frac{\kappa}{\Omega_{k}}\right)^{-2} \left( \ddlnr{\tau} + \m{\tilde\tau}\ddlnr{\ln{\left[r\mathscr{R}\right]}} \right) > 1 ~.
\end{equation}
It is important to note that $\kappa$ contains dependences on both radiation and gas pressure. In the interest of specifically studying IRI, we consider the case when the rotation curve is solely modified by radiation pressure. Then $\kappa$ can be expressed as
\begin{equation}
\label{eq:sim_kappa}
\left(\frac{\kappa}{\Omega_{k}}\right)^{2} = 1 - \Bo e^{-\tau}\ddlnr{\ln{\left[r\Bo\right]}} + \Bo e^{-\tau} \ddlnr{\tau} ~.
\end{equation}
Plugging \eqnref{eq:sim_kappa} into \eqnref{eq:disper}, the condition for instability becomes
\begin{equation}
\label{eq:crit2}
q_{\beta} \equiv \Bo e^{-\tau} \left( \ddlnr{\ln{\left[r\Bo\right]}} + \m{\tilde\tau} \ddlnr{\ln{\left[r\mathscr{R}\right]}} \right) > 1 ~.
\end{equation}

To complete our derivation, we need to evaluate $\m{\tilde\tau}$. We begin by integrating \eqnref{eq:tau_m} by parts
\begin{equation}
\label{eq:tau_m_int}
\m{\tilde\tau} = \tau - ik_{r} \frac{\cs^2}{\m{\eta}} \int^r_0 \frac{\m{\eta}}{\cs^2} \tau \dr' ~.
\end{equation}

If in the disk there exists a ''transition region'' where the disk sharply transitions from being radially transparent to opaque, then one can show that inside this region, the second term on the right-hand side of \eqnref{eq:tau_m_int} has a magnitude of the order $k_{r} \Delta r$, where $\Delta r$ is the width of the transition region. This allows us to approximate $\m{\tilde\tau} \sim \tau$ in the limit $\frac{1}{\Delta r}\gg k_{r}$. Moreover, even when $\frac{1}{\Delta r}\sim k_{r}$, we expect $\m{\tilde\tau} \sim \tau$ to remain accurate to within the order of unity.
In \secref{sec:tau_m} we evaluate $\m{\tilde\tau}$ explicitly and find out to what extend this holds true.

While \eqnref{eq:crit1} is the more general form, \eqnref{eq:crit2} does reveal surprising behavior: it contains no explicit dependence on $\ddr{\tau}$, as it is completely canceled by the stabilizing effect of $\kappa^2$. Replacing it is a term containing $\ddr{\beta}$, whose effect is to lower $\kappa^2$ to the point of triggering a form of irradiation-induced Rayleigh instability. While it does contribute to the instability of the disk, we do not consider it the true trigger of IRI. Rather, we focus on the second term inside the bracket. First, it implies that a disk is unstable to IRI if it has a positive gradient in $\mathscr{R}$, which can be created by a gradient in $\Sigma$ and/or $\Bo$. Second, this gradient must be located where $\m{\tilde\tau}e^{-\tau}\sim \tau e^{-\tau}$ is reasonably large, which is precisely the transition region. This is consistent with our picture that IRI is driven by shadowing. Because of the uncertainty in $\m{\tilde\tau}$, as well as the other assumptions stated in the beginning of this section, Equations \ref{eq:crit1} and \ref{eq:crit2} should be taken as order-of-magnitude guidelines rather than rigid conditions.

%==================================================================
\subsection{Corotating Modes}
\label{sec:cor}
%==================================================================
If a linear mode exists, its corotation radius can be found by solving \eqnref{eq:2ndeq} for $\m{\Omega}=0$. Similar to \secref{sec:crit}, we apply the WKB approximation, and then the real part of \eqnref{eq:2ndeq} evaluated at the corotation radius can be rewritten as:
\begin{equation}
\label{eq:cor1}
\m{\Omega} = 0 = 2m\Omega \frac{\frac{h^2}{r^2} \ddlnr{\ln{\mathscr{F}}} - \Bo e^{-\tau}\m{\tilde\tau}}{ \frac{\kappa^2}{\Omega_k^2} + |k|^2 h^2 - \Bo e^{-\tau} \left( \ddlnr{\tau} + \m{\tilde\tau} \ddlnr{\ln{\left[r\mathscr{R}\right]}} \right) } ~,
\end{equation}
where $|k|^2 = k_r^2 + m^2 / r^2$, and $\mathscr{F} \equiv \Sigma \Omega / \kappa^2$ is a quantity inversely proportional to the vortensity of the disk. The corotation radius is therefore located at where the following condition is satisfied:
\begin{equation}
\label{eq:cor2}
\ddlnr{\ln{\mathscr{F}}} = \left(\frac{h}{r}\right)^{-2}\Bo e^{-\tau}\m{\tilde\tau} ~.
\end{equation}
For barotropic flow and $\Bo=0$, this condition becomes identical to that described in Section 2.2 of \citet{RWI1} for RWI. The usefulness of \eqnref{eq:cor2} is limited because without a full solution, the exact value of $\m{\tilde\tau}$ is unknown. However, allowing that $\m{\tilde\tau} \sim \tau$, it does provide an insight:  since the the right-hand side of \eqnref{eq:cor2} is always positive, if $\mathscr{F}$ contains a local maximum, the corotation radius will always be located at a lower orbit than where this maximum is. In our disk model described in the following section, $\mathscr{F}$ does contain a local maximum within the transition region, so we expect the corotation radius to be smaller for disks with a larger value of $\left(\frac{h}{r}\right)^{-2}\Bo$. This prediction is tested in \secref{sec:linear}.

%==================================================================
\section{Disk Model}
\label{sec:disk}
\begin{figure}[]
\includegraphics[width=0.99\columnwidth]{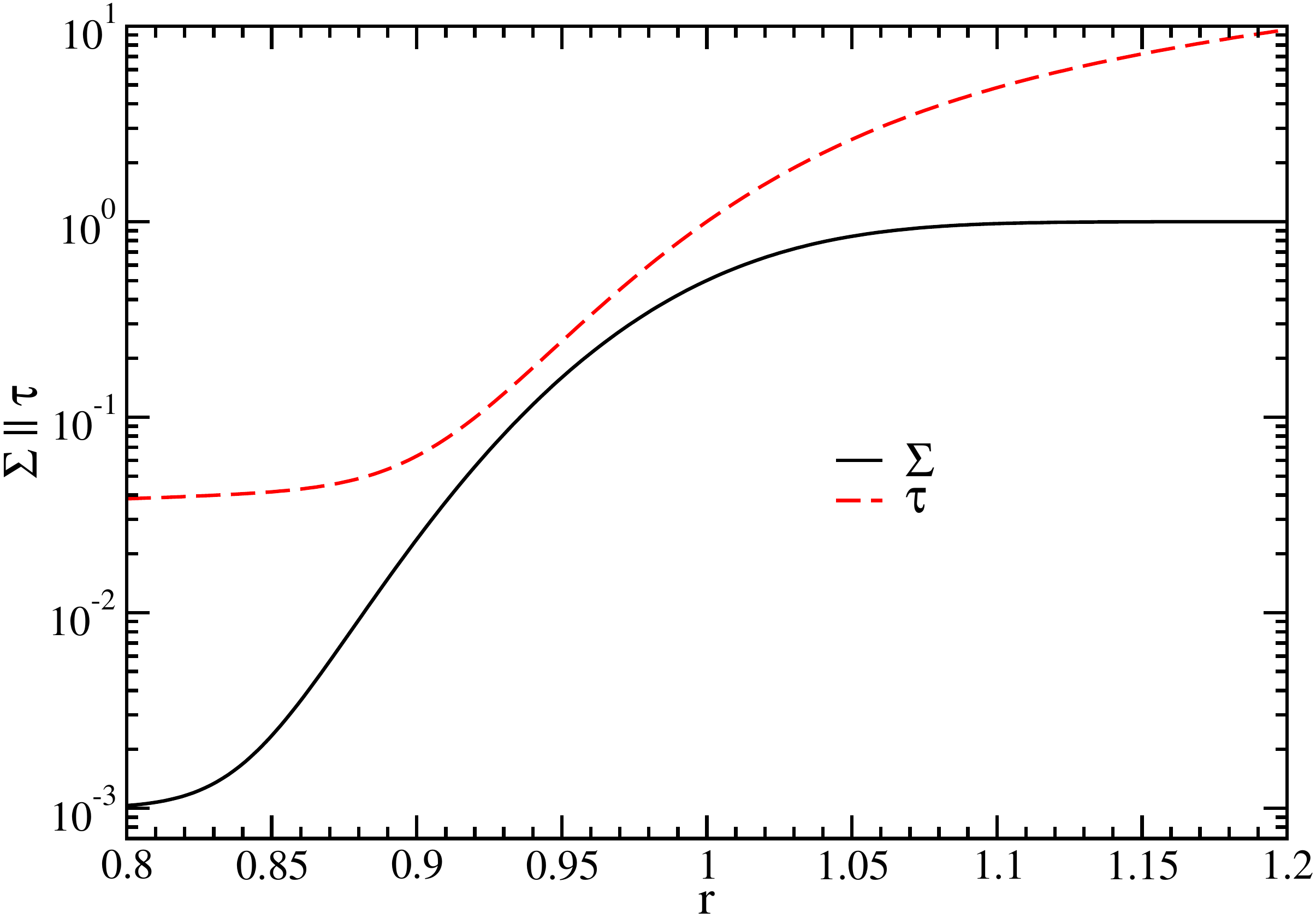}
\caption{Black solid line plotting the surface density profile described by \eqnref{eq:sden} and red dashed line plotting the optical depth profile.}
\label{fig:sden}
\end{figure}

%==================================================================
For simplicity we do not consider any spatial variation in the composition of the disk, therefore $\Bo$ and $\kappa_{\rm opa}$ are constants. With this simplification, \eqnref{eq:crit2} says that the disk is most unstable if $\mathscr{R}$ has a large positive gradient near $\tau=1$. We create this condition with a disk that contains a sharp inner edge. At this edge, $\Sigma$ increases by orders of magnitude across a small radial range, while $\tau$ rises from a small value to above unity. Our prescription for such a disk is:
\begin{equation}
\label{eq:sden}
\Sigma(r) = \frac{1}{2}\left(\Sigma_{d}+\Sigma_{\rm c}\right)+\frac{1}{2}{\rm erf}\left(\frac{r-r_0}{\sqrt{2{\Dr}^2}}\right)\left(\Sigma_{d}-\Sigma_{\rm c}\right),
\end{equation}
where $\Sigma_{d}$ is the surface density of the disk, $\Sigma_{c}$ is the surface density inside the cavity , $r_0$ is the radius at which the inner edge is located, and $\Delta r$ is the width of this edge. We set $\Sigma_{d}=1$ and $\Sigma_{c}=0.001$ for a density contrast of $10^3$. We also set $r_0=1$ and $GM=1$ so that the dynamical time $t_{\rm dyn}$ at the edge is $\Omega_{k}^{-1}=1$. For the sharpness of the edge, we set $\Delta r = 0.05$. The motivation for this choice is that $\Delta r$ is unlikely to be shorter than $h$, which for protoplanetary disks has a typical value of $0.05r$. $\kappa_{\rm opa}$ is chosen such that $\tau(r_0)=1$. If we move this $\tau=1$ point to a much smaller/larger radius, the disk edge will be become optically thick/thin, and thus one would expect the instability to weaken or even disappear. \figref{fig:sden} plots both the $\Sigma$ and $\tau$ profile. To complete the equation set, we adopt an isothermal equation of state so that $\cs$ is a constant.

This leaves two free parameters in our model: $\Bo$ and $\cs$. We perform a parameter study over the range $\Bo=\{0,0.3\}$ and $\cs=\{0.02,0.06\}$. Note that for $h(r_0)\gtrsim\Delta r$, corresponding to $\cs\gtrsim0.05$, the disk edge may become hydrodynamically unstable. We deliberately include this limit in our parameter space both as a sanity check and to investigate how IRI can be differentiated from other forms of instabilities.

%==================================================================
\section{Two Independent Approaches}
\label{sec:two}
%==================================================================
For our given disk model, we aim to find out for the IRI (1) how the modal growth rate varies as a function of $\Bo$ and $\cs$, and (2) what are the properties of its nonlinear phase. Two independent approaches are used: a numerical method using hydrodynamical simulations and a semi-analytic method that solves the linearized problem (\eqnref{eq:2ndeq}). These two methods not only serve as verifications for each other, but are also complementary since a full simulation gives us an insight into the nonlinear phase, while the semi-analytic method is not subjected to limitations such as resolution and numerical noise.

%==================================================================
\subsection{Hydrodynamical Simulation}
\label{sec:sim}
%==================================================================
We numerically simulate the 2D disk described in \secref{sec:disk}. The code we use is the Lagrangian, dimensionally-split, shock-capturing hydrodynamics code \texttt{PEnGUIn} ({\bf P}iecewise Parabolic Hydro-code {\bf En}hanced with {\bf G}raphics Processing {\bf U}nit {\bf I}mplementatio{\bf n}), which has been previously used to simulate disk gaps opened by massive planets \citep{Fung2014}. It uses the piecewise parabolic method (PPM; \citealt{PPM}), and its main solver is modeled after VH-1 \citep{VH1}, with a few of the same modifications as described in \citet{Fung2014}. It solves \eqnref{eq:cont} and \ref{eq:moment1}, and contains an additional module to compute $\tau$ using piecewise parabolic interpolation to match the order of PPM.

Our simulations have a domain spanning 0.5 to 2.0 in radial (in units where the disk edge is located at $r_0=1$) and the full 0 to $2\pi$ in azimuth. Moving the inner boundary to 0.7 or the outer boundary to 1.5 has a negligible effect on the growth of linear modes. We opt for a larger domain to accommodate the more violent nonlinear evolution.

\begin{figure}[]
\includegraphics[width=0.99\columnwidth]{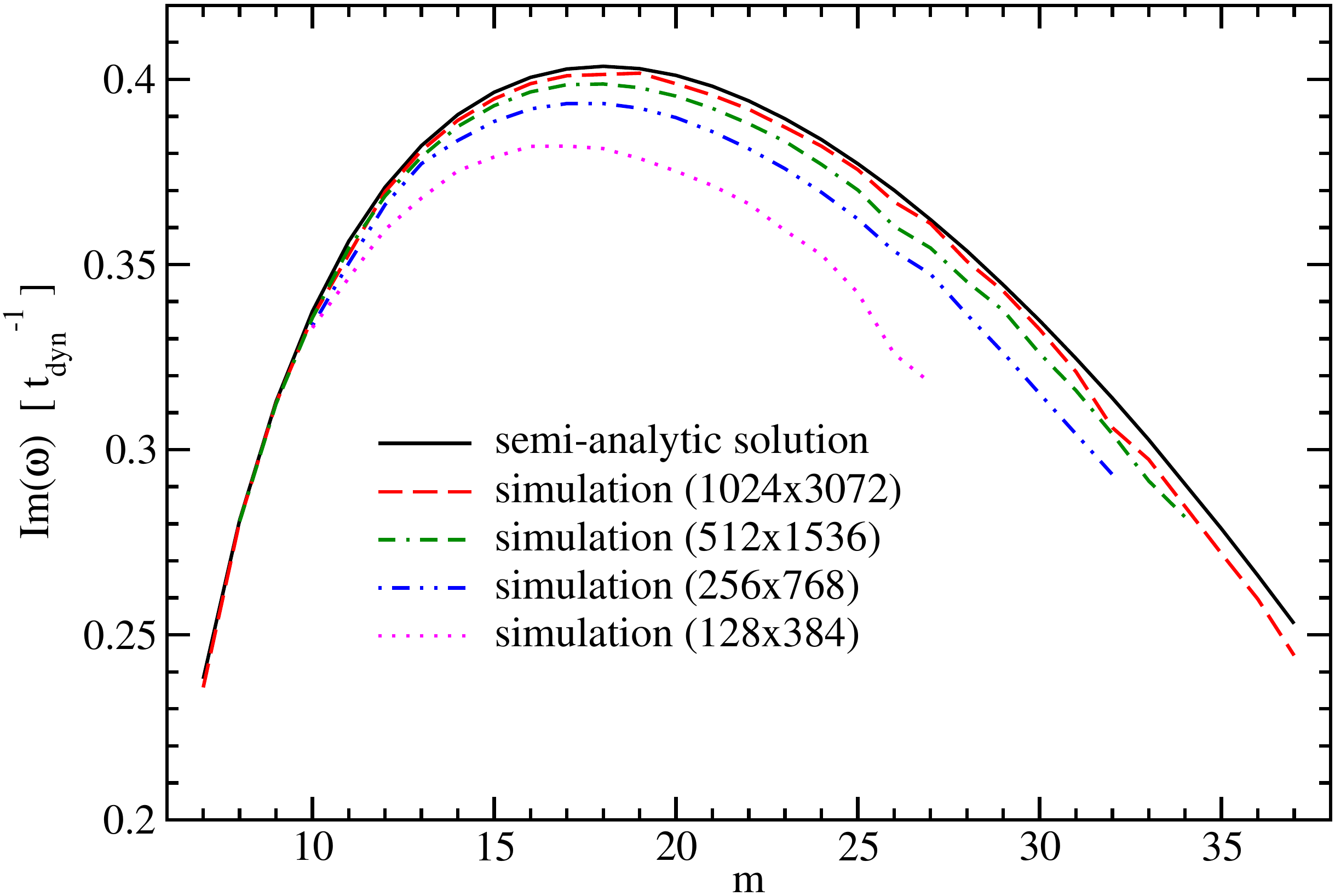}
\caption{Growth rates of azimuthal modes with $(\Bo, \cs)$ = $(0.2, 0.02)$. At 1024 (r) by 3072 ($\phi$), the growth rates extracted from simulation match those found by the semi-analytic method to $\sim1\%$. For this particular case, the fastest growing mode is $m=18$, with a growth rate of ${\rm Im}(\omega)=4.0\times10^{-1} t_{\rm dyn}^{-1}$. See \secref{sec:linear} for further discussions on how these results vary with $\Bo$ and $\cs$.}
\label{fig:agree}
\end{figure}

The resolution is 1024 (r) by 3072 ($\phi$). Azimuthal grid spacing is uniform everywhere, but radial grid spacing is uniform only between 0.5 and 1.3; from 1.3 to 2.0 it is logarithmic. This takes advantage of \texttt{PEnGUIn}'s ability to utilize non-uniform grids to enhance the resolution around the disk edge. The resulting grid size at $r_0$ is about 0.001 (r) by 0.002 ($\phi$). This gives at least 10 cells per $h$ for even the smallest $h$ we consider. \figref{fig:agree} shows how our simulations converge with resolution.

\begin{figure}[]
\includegraphics[width=0.99\columnwidth]{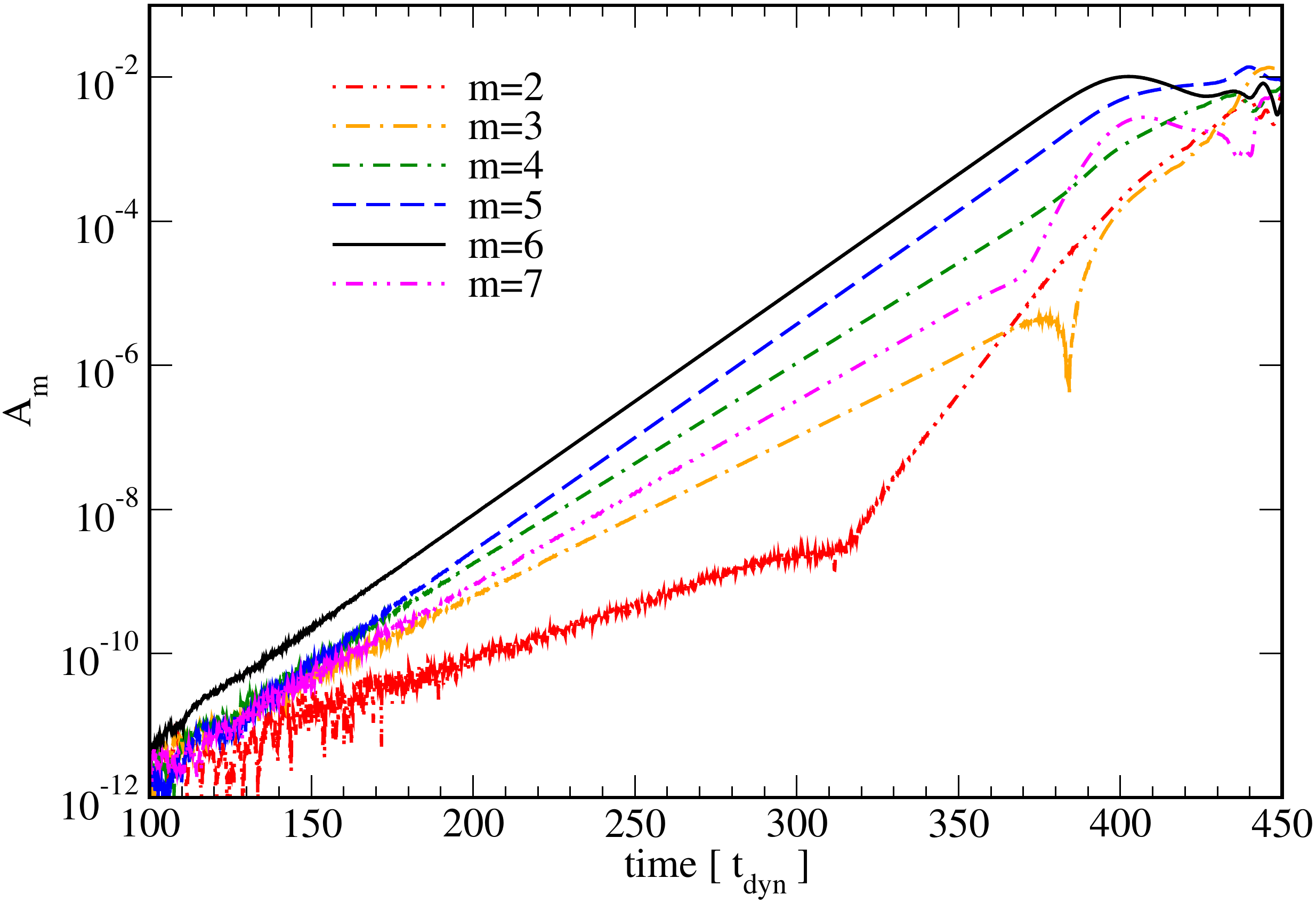}
\caption{Temporal evolution of $\m{A}$ (see \eqnref{eq:A_m}) with $(\Bo, \cs)$ = $(0.05, 0.05)$. A well-defined exponentially growing phase can be seen around $t = 200\sim300$. Beyond $t=300$ the modes begin to exhibit higher-order coupling.}
\label{fig:time}
\end{figure}

In each simulation, we extract the amplitudes of azimuthal modes as functions of time, resolving up to $m=50$:
\begin{equation}
\m{A}(t) = \frac{1}{2\pi}\left|\int^{2\pi}_{0}\int^{1.1}_{1.0}\Sigma(t) e^{im\phi} {\rm d}r~ {\rm d}\phi \right| ~,
\label{eq:A_m}
\end{equation}
where we have chosen to integrate over the radial range $r=\{1.0,1.1\}$. Instantaneous values of $\m{A}$ are not the focus; rather, we seek a distinct period of exponential growth where we can measure its growth rate, i.e., the imaginary part of $\omega$. \figref{fig:time} shows one example of how modal growth behaves in these simulations. For a disk of a given set of parameters, the highest growth rate characterizes its timescale for instability.

We use a boundary condition fixed to the initial values described by Equations \ref{eq:omg} and \ref{eq:sden}, with zero radial velocity. To reduce noise in $\m{A}$, we also include wave-killing zones in $r=\{0.5,0.6\}$ for the inner boundary and $r=\{1.6,2.0\}$ for the outer. Within these zones, we include an artificial damping term:
\begin{equation}
\frac{\partial{X}}{\partial t} = \left(X(t=0)-X\right)\frac{2\cs|r-r_{\rm kill}|}{d_{\rm kill}^2} ~,
\end{equation}
where $X$ includes all disk variables $\Sigma$, $P$, and $\bf v$; $r_{\rm kill}$ is the starting radius of the wave-killing zone, which is 0.6 for the inner boundary and 1.6 for the outer; and $d_{\rm kill}$ is the width of these zones, which equals 0.1 for the inner boundary and 0.4 for the outer. In the end we are able to resolve $\m{A}$ as small as $10^{-10}$, such as shown in \figref{fig:time}.

Simulations are terminated soon after the instability becomes fully nonlinear: up to 100 orbits, or 628 $t_{\rm dyn}$. For very slowly growing modes, numerical noise severely hampers the precision of growth rate measurements. Consequently, this method is only capable of measuring growth rates $\gtrsim 0.01 t_{\rm dyn}^{-1}$. The computational time for \texttt{PEnGUIn} is about 12 minutes per orbit on a single GTX-Titan graphics card.

%==================================================================
\subsection{Semi-analytic Method}
\label{sec:ana}
%==================================================================
\eqnref{eq:2ndeq} constitutes an eigenvalue problem, where $\m{\eta}$ is the eigenfunction and $\omega$ is the eigenvalue. To solve this problem, we develop a code that directly integrates the differential equations, iterates for the correct boundary conditions, and optimizes to find the eigenvalues. The complexity of this code is mainly to overcome the difficulty imposed by the integral in \eqnref{eq:2ndeq}, which effectively raises the order of the differential equation. The details are documented in Appendix \ref{append}.

Despite the fact that it solves the linearized equations, our semi-analytic method in fact requires a much longer computational time than simulations using \texttt{PEnGUIn}. Due to limited resources, initially we only apply it to five sets of parameters. \tabref{tab:ana} contains a list of these sets. One major advantage of this method is that it does not have a limit to how slow of a growth rate can be detected, so we also apply it to all cases where simulations do not detect any modal growth. Among them, we find a positive growth rate for one case. \footnote{This case has $(\Bo, \cs) = (0, 0.05)$. Since $\Bo=0$, the modal growth is purely hydrodynamical and unrelated to IRI. See \secref{sec:linear} for further discussions.} It is also listed in \tabref{tab:ana}, making a total of six sets of parameters.

\begin{deluxetable}{lcccc} 
\tablecolumns{5} 
\tablewidth{0pc} 
\tablecaption{Semi-analytic Results\tablenotemark{a}} 
\tablehead{
\colhead{$\Bo$} & 
\colhead{$\cs$} & 
\colhead{Im($\omega$) $(t_{\rm dyn}^{-1})$} & 
\colhead{$m$} &
\colhead{$r_{\rm cor}$\tablenotemark{b} $(r_0)$}
}
\startdata 
0    & 0.06 & $7.7\times10^{-2}$ & 4  & 1.046\\
0    & 0.05 & $2.5\times10^{-3}$ & 1  & 1.052\\
0.1  & 0.05 & $7.2\times10^{-2}$ & 6  & 0.979\\
0.15 & 0.04 & $1.6\times10^{-2}$ & 7  & 0.956\\
0.15 & 0.03 & $1.4\times10^{-1}$ & 8  & 0.943\\
0.2  & 0.02 & $4.0\times10^{-1}$ & 18 & 0.938
\enddata
\tablenotetext{a}{We only report the properties of the fastest growing mode.}
\tablenotetext{b}{$r_{\rm cor}$ denotes the corotation radius.}
\label{tab:ana}
\end{deluxetable}

%==================================================================
\section{Results}
\label{sec:result}
%==================================================================

\begin{figure*}[]
\includegraphics[width=1.99\columnwidth]{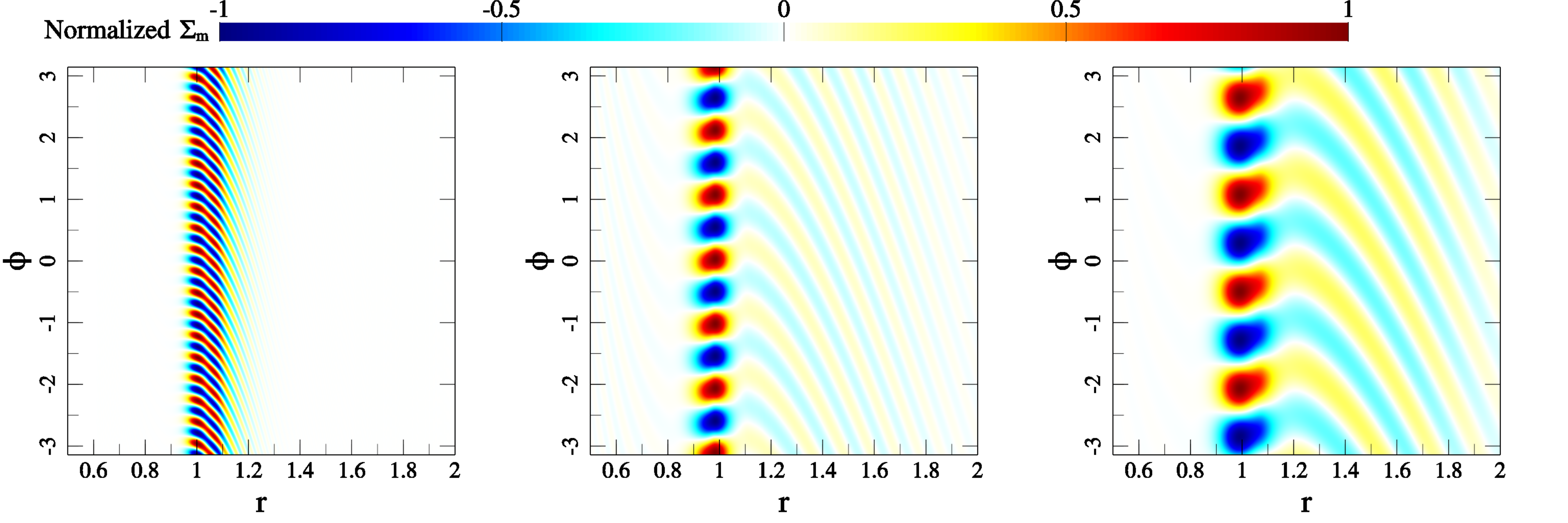}
\caption{Fastest growing modes extracted from simulations through Fourier decomposition. Color shows the surface density normalized to the peak of each mode. On the left is an $m=18$ mode from $(\Bo,\cs)=(0.2,0.02)$; in the middle is $m=6$ from $(\Bo, \cs)=(0.1,0.05)$; and on the right is $m=4$ from $(\Bo, \cs)=(0,0.06)$.}
\label{fig:sim_mod}
\end{figure*}

\begin{figure*}[]
\includegraphics[width=1.99\columnwidth]{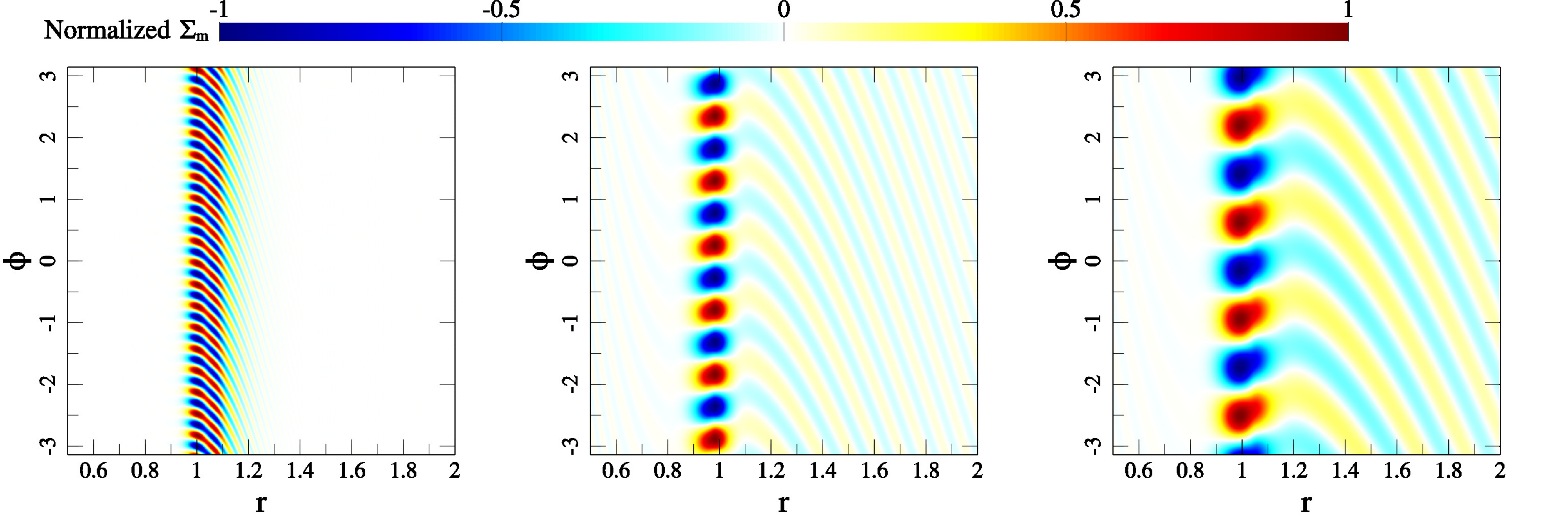}
\caption{The fastest growing modes directly computed using our semi-analytic method for the same parameters listed in \figref{fig:sim_mod}.}
\label{fig:ana_mod}
\end{figure*}

The sets of parameters we consider are $\Bo=\{0, 0.05, 0.1, 0.15, 0.2, 0.25, 0.3\}$ and $\cs=\{0.02, 0.03, 0.04, 0.05, 0.06\}$. All 30 combinations of these values are simulated, but only a select few are solved with our semi-analytic method (see \tabref{tab:ana}). The growth rates found by our two independent approaches agree to $\sim1\%$. \figref{fig:agree} gives one example of this agreement. Also, the shapes of the modes extracted from simulations are nearly identical to the ones solved semi-analytically. Comparing \figref{fig:sim_mod} to \ref{fig:ana_mod}, results from the two methods are only distinguishable near the outer boundary, where the simulated ones show some artificial damping due to the wave-killing zones imposed. Because of the excellent agreement we are able to combine the results of the two approaches to give a detailed picture for the IRI linear modes, complemented by the nonlinear evolution provided by simulations.

%==================================================================
\subsection{Linear Modes}
\label{sec:linear}
%==================================================================

We find clear growth of asymmetric modes for all cases with $\Bo$ larger than a certain threshold value that is weakly dependent on $\cs$ across our parameter space. For most of our chosen $\cs$ values, modal growth is only detected when $\Bo\geq0.1$, except for $\cs\sim0.02$, where this threshold rises to $\Bo\geq0.15$. From a simple perspective, we expect the disk to be more unstable for a larger $\Bo$ and smaller $\cs$, because $\Bo$ measures the strength of radiation pressure while $\cs$ is a source of resistance to external forcing. In general, we do find the growth rate to increase with $\Bo$ and decrease with $\cs$, but with obvious exceptions.

\begin{figure}[]
\includegraphics[width=0.99\columnwidth]{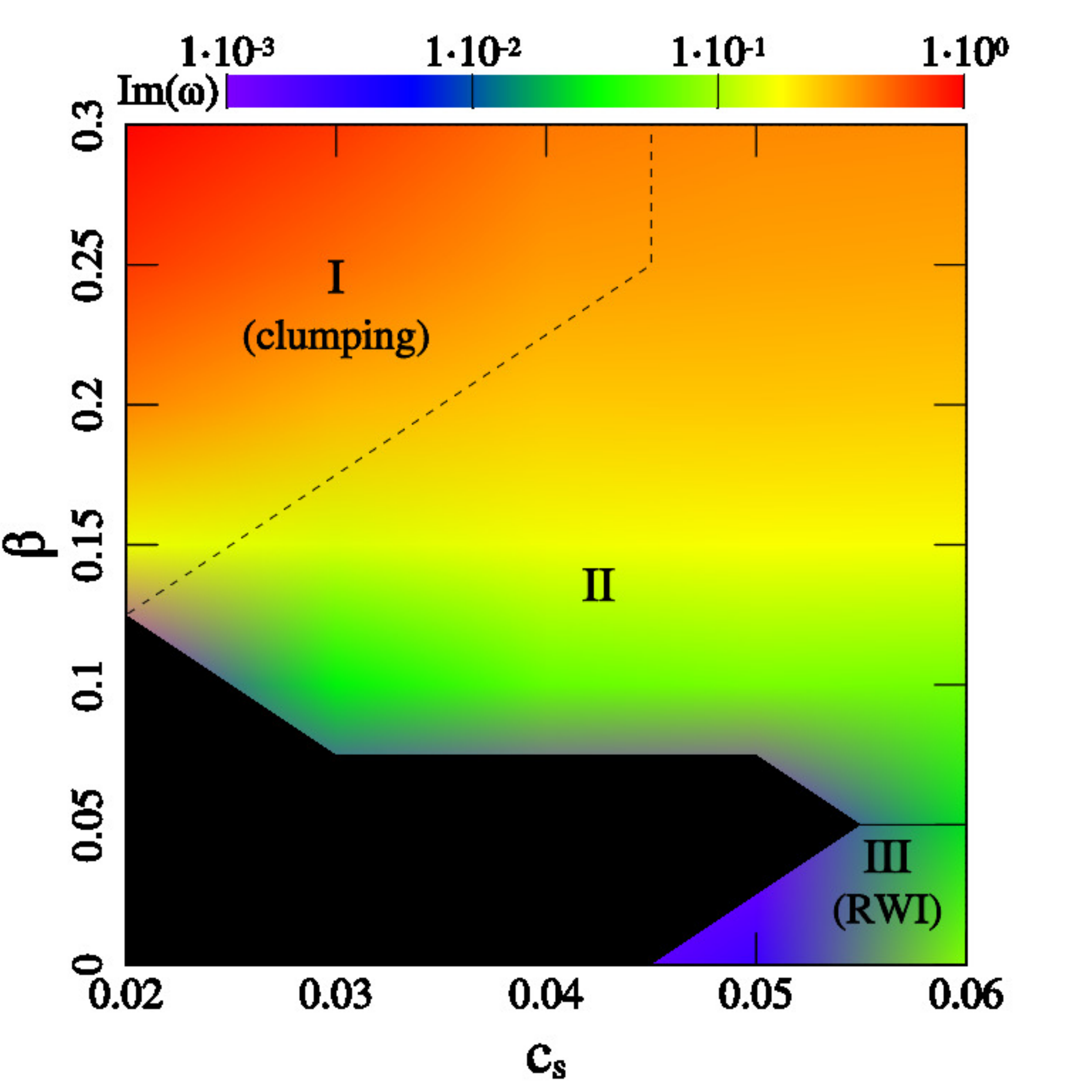}
\caption{Growth rate of the fastest growing mode as a function of $\Bo$ and $\cs$. The black region is where a positive growth rate is not found with both of our approaches. Regions I and II are where IRI operates, while region III sees the purely hydrodynamical RWI. In the nonlinear phase, clumping occurs in region I, where local surface density is enhanced by at least a factor of two, often much higher.}
\label{fig:para}
\end{figure}

\begin{figure}[]
\includegraphics[width=0.99\columnwidth]{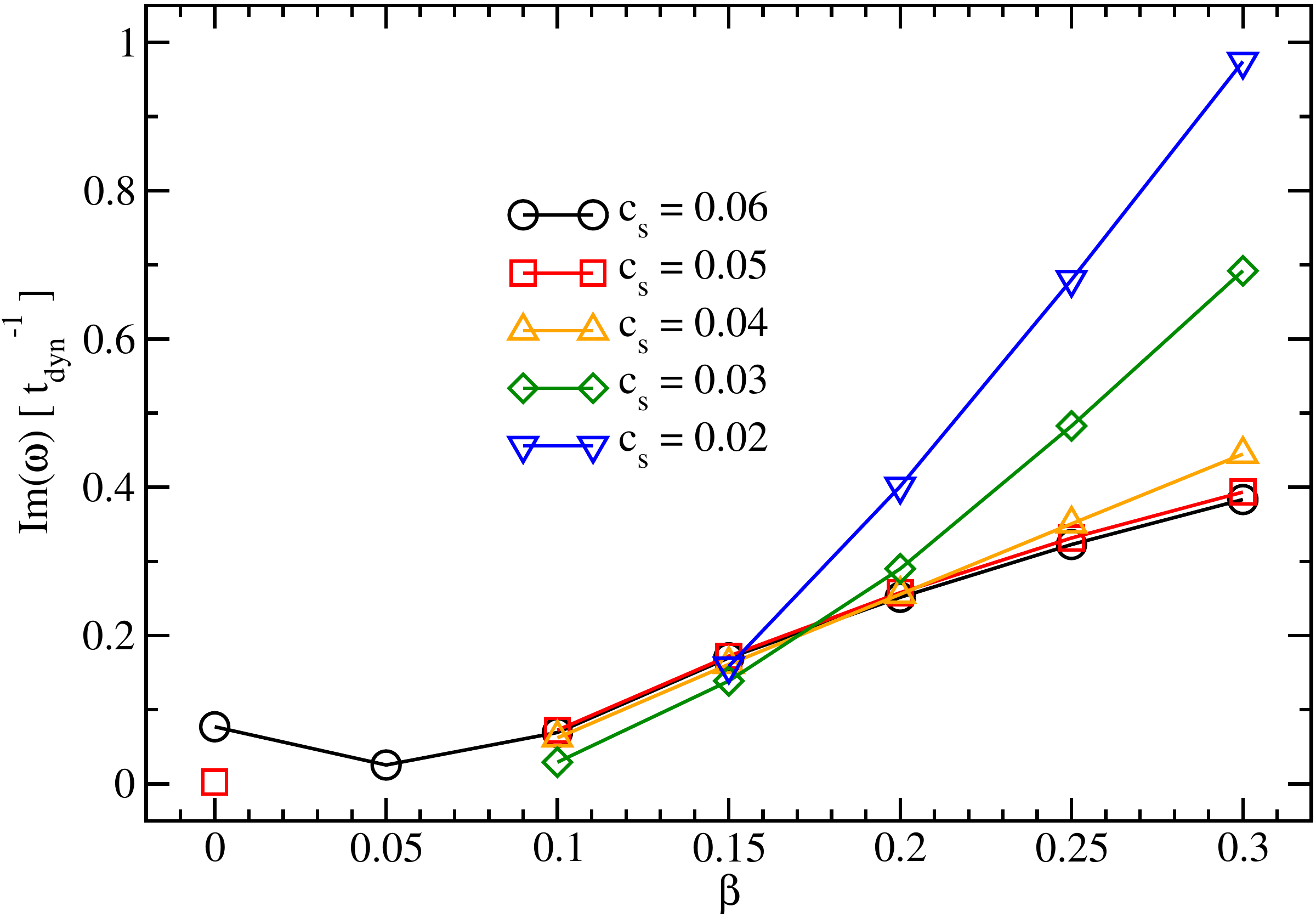}
\caption{Growth rate of the fastest growing mode as a function of $\Bo$. The $(\Bo, \cs)=(0,0.05)$ point is disconnected because no modal growth is detected at $(\Bo, \cs)=(0.05,0.05)$.}
\label{fig:Bo_max}
\end{figure}

In \figref{fig:para} we divide our parameter space into three regions: regions I and II where modal growth is driven by radiation pressure, and region III where it is mainly driven by
hydrodynamical effects.
In regions I and II, growth rate scales roughly linearly with $\Bo$ for any given $\cs$, a trend that can be more easily seen in \figref{fig:Bo_max}. This is consistent with our expectations. 

\begin{figure}[]
\includegraphics[width=0.99\columnwidth]{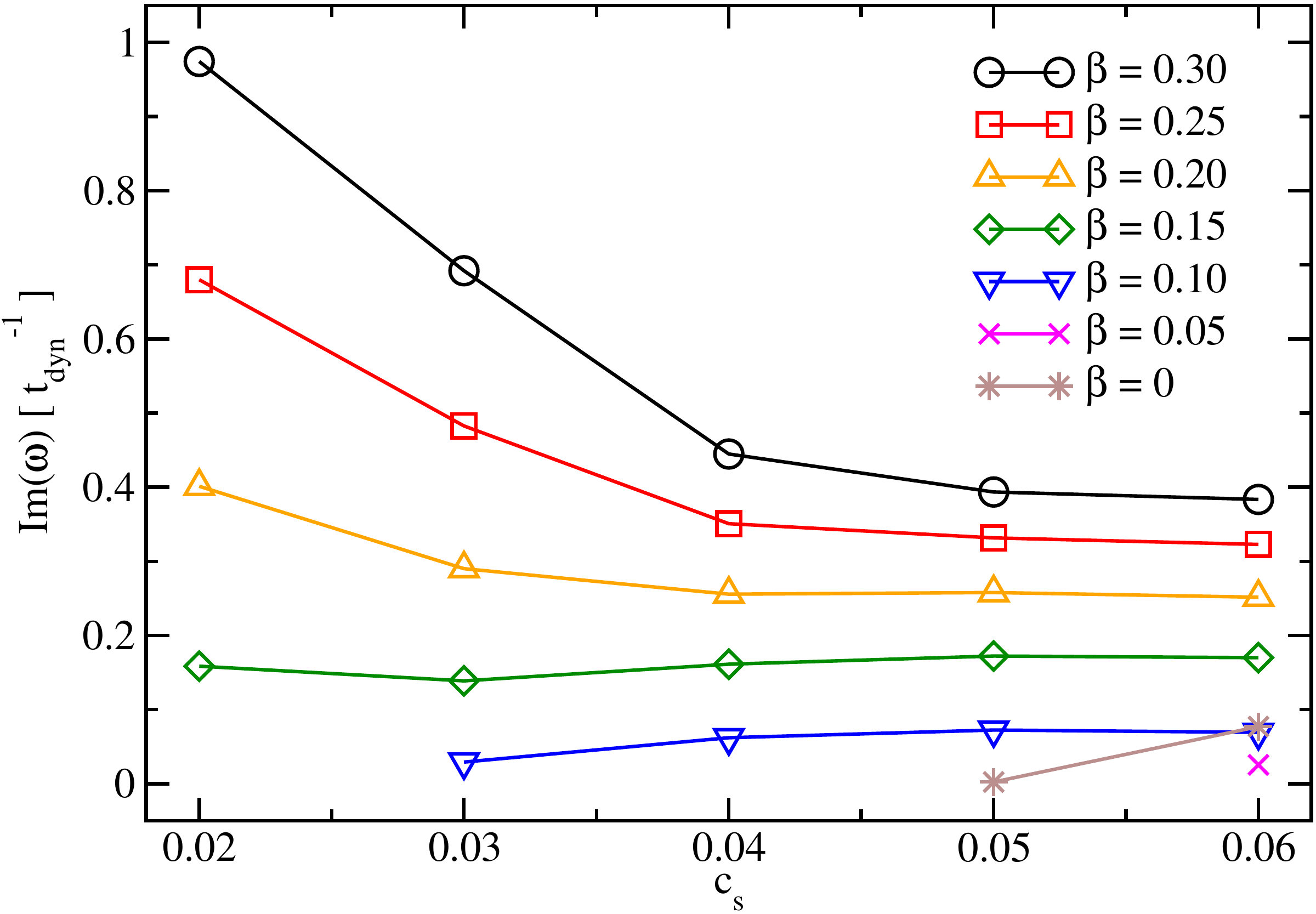}
\caption{Growth rate of the fastest growing mode as a function of $\cs$.}
\label{fig:cs_max}
\end{figure}

\figref{fig:cs_max}, however, reveals a more complicated aspect of IRI. Disregarding the $\Bo=0$ data points that belong to region III, the growth rate is very close to constant over the range $0.04\leq \cs \leq 0.06$ for any given $\Bo$. Once $\cs$ goes below $0.04$ it shows different trends depending on the value of $\Bo$: the growth rate increases as $\cs$ decreases for $\Bo\geq0.2$, but for a smaller $\Bo$ the trend flattens or even begins to drop. This complex behavior may relate to how sound waves and IRI modes couple. While sound waves have a length scale $h$, IRI modes are mainly restricted by the sharpness of the transition region, which has a length scale $\Delta r$. Our results suggest that the coupling is weak when $h\sim\Delta r$, and becomes much stronger as $h$ decreases, allowing the transition region to accommodate the full wavelength of the longest wave.

Region III is where radiation pressure becomes a smaller effect than gas pressure. For $\cs \geq 0.05$, or equivalently, $h(r_0)\geq\Delta r$, we detect modal growth even in the absence of any radiation pressure. 
In fact, when $\Bo=0$, our disk model is similar to the ''homentropic step jump'' model used by \citet{RWI2} to study RWI (see their Figure 2). Their Figure 11 shows that for a pressure jump with a width $\Delta r=0.05$, RWI modes will develop if $\cs\gtrsim0.06$ \footnote{There are a few small differences between the disk model used by \citet{RWI2} and ours. For example, they use an adiabatic equation of state with an adiabatic index of $5/3$, and their pressure jump is modeled with a different formula (compare their Equation 3 to our \eqnref{eq:sden}). We consider these differences insignificant.}. Our results are consistent with their findings.

\begin{figure}[]
\includegraphics[width=0.99\columnwidth]{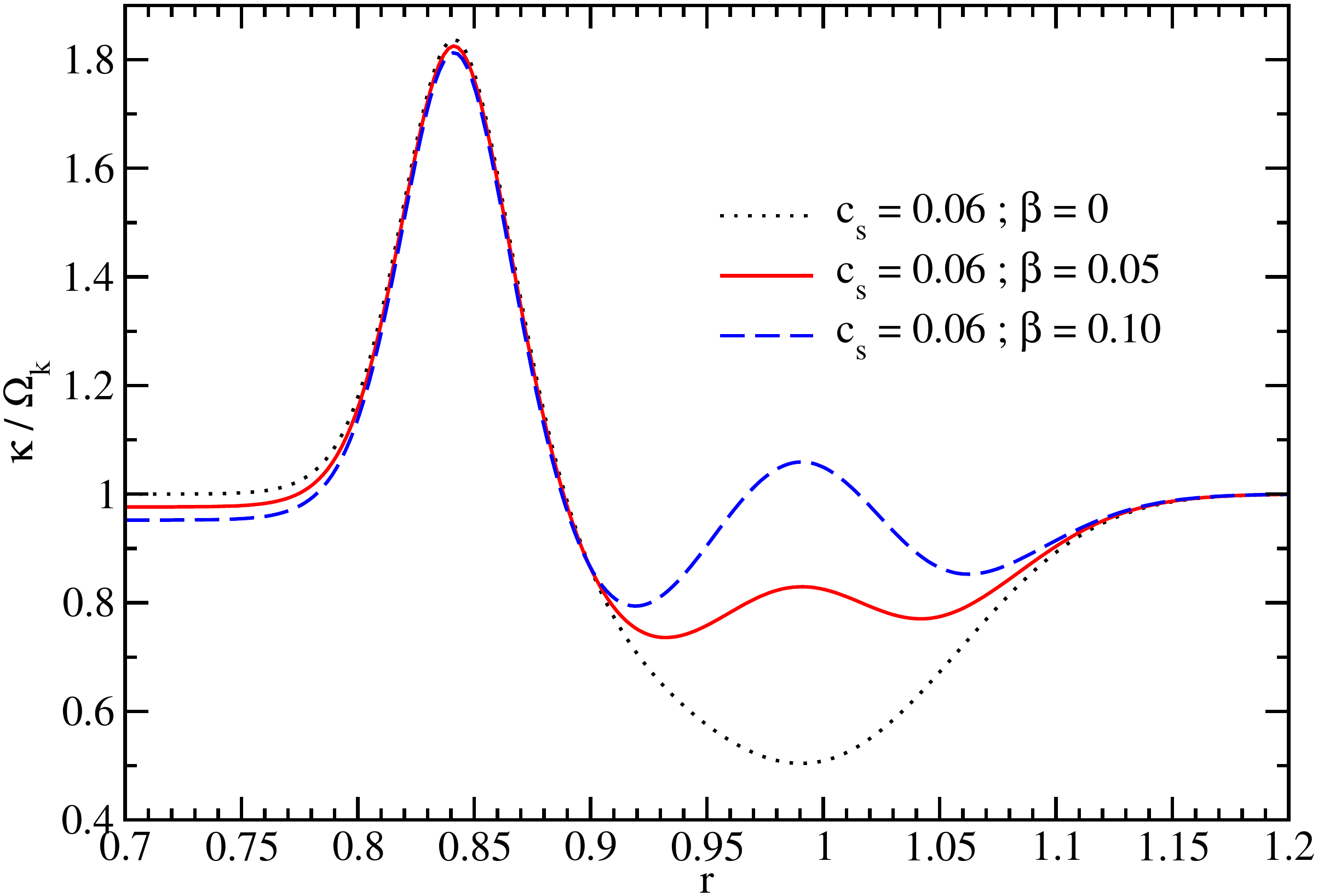}
\caption{$\kappa$ vs. $r$ for three sets of parameters. The black dotted curve shows $\kappa$ modified by gas pressure only. As $\Bo$ increases, the local minimum near $r=1$ is flattened (red solid curve) and reversed (blue dashed curve).}
\label{fig:kappa}
\end{figure}

The division between IRI and RWI is clear to us because the two mechanisms appear to destructively interfere with each other. For $\cs = 0.06$, there is a clear drop in growth rate from $\Bo=0$ to $\Bo=0.05$ before it rises again (see \figref{fig:Bo_max}). Similarly for $\cs = 0.05$, we do not detect any modal growth at $\Bo=0.05$ even though it is detected at both $\Bo=0$ and $\Bo=0.1$. One clue to this behavior is that we find $\cs$ and $\Bo$ to have opposing effects on the epicyclic frequency $\kappa$. In \figref{fig:kappa} we see that gas pressure lowers $\kappa$ near $r=r_0$, while $\Bo$ raises it. Two effects roughly cancel when $\Bo=0.05$. It is unclear whether this is coincidental or not. 

This dividing line may not remain at $\Bo=0.05$ for a different value of $\Delta r$ or $\cs$. If we create a sharper edge by reducing $\Delta r$, both IRI and RWI are expected to be enhanced and it is unclear to us whether this dividing line will move to a higher or lower $\Bo$. Additionally, a high $\cs$ can push $\kappa^2$ below zero and trigger Rayleigh instability which further complicates the matter. For our disk model this limit is at $\cs\sim0.07$. Since the focus of this paper is to characterize IRI, we defer the thorough investigation on the interactions between IRI and other forms of instability to a future study.

Other than the growth rate, we also find other general trends about the linear modes. For an increasing $\Bo$ or decreasing $\cs$,
the azimuthal mode number $m$ of the fastest growing mode increases. The dependence on $\Bo$ is particularly pronounced. In the most extreme case, the fastest growing mode is $m=47$ when $(\Bo, \cs)=(0.3,0.02)$. In contrast, for all cases with $\Bo=0.1$, $m=5\sim6$ is the fastest growing mode. \figref{fig:agree} shows an intermediate case where the fastest growing mode is $m=18$. See \tabref{tab:ana} for more examples. Similarly, we find that the radial extent of the fastest growing mode becomes more confined as $\Bo$ increases and $\cs$ decreases, as can be seen in Figures \ref{fig:sim_mod} and \ref{fig:ana_mod}. It is therefore empirically apparent that a higher $\Bo$ encourages the growth of a shorter wavelength mode. While our theory does not make any predictions about which mode grows the fastest, in hindsight this result is not surprising because the radial motion of an IRI mode must be driven by radiation pressure, so a stronger perturbing force should generate a faster radial motion, and therefore a higher frequency wave.

Another trend is that for an increasing $\Bo$ or decreasing $\cs$, the corotation radius decreases. This is in accordance with our prediction in \secref{sec:cor}. The dependence is weak but noticeable (see \tabref{tab:ana}). Curiously, \figref{fig:sim_mod} and \ref{fig:ana_mod} show that the peak location of each mode is relatively insensitive to variations in both $\Bo$ and $\cs$. Consequently, these peaks generally do not coincide with their corotation radii.

For the bulk of this work we do not explicitly vary $\Delta r$ as a free parameter. Since our choice of $\Delta r=0.05$ is arbitrary, it is useful to find out to what extent our results would change for a different $\Delta r$. Setting $\cs=0.04$ and $\Delta r=0.1$, we find the threshold for modal growth becomes $\Bo\geq0.25$, roughly twice as large as when $\Delta r=0.05$. This suggests that the threshold value scales roughly linearly with $\Delta r$ for the parameter space we considered.

%==================================================================
\subsection{Nonlinear Evolution}
\label{sec:nonlinear}
%==================================================================

\begin{figure*}[]
\includegraphics[width=1.99\columnwidth]{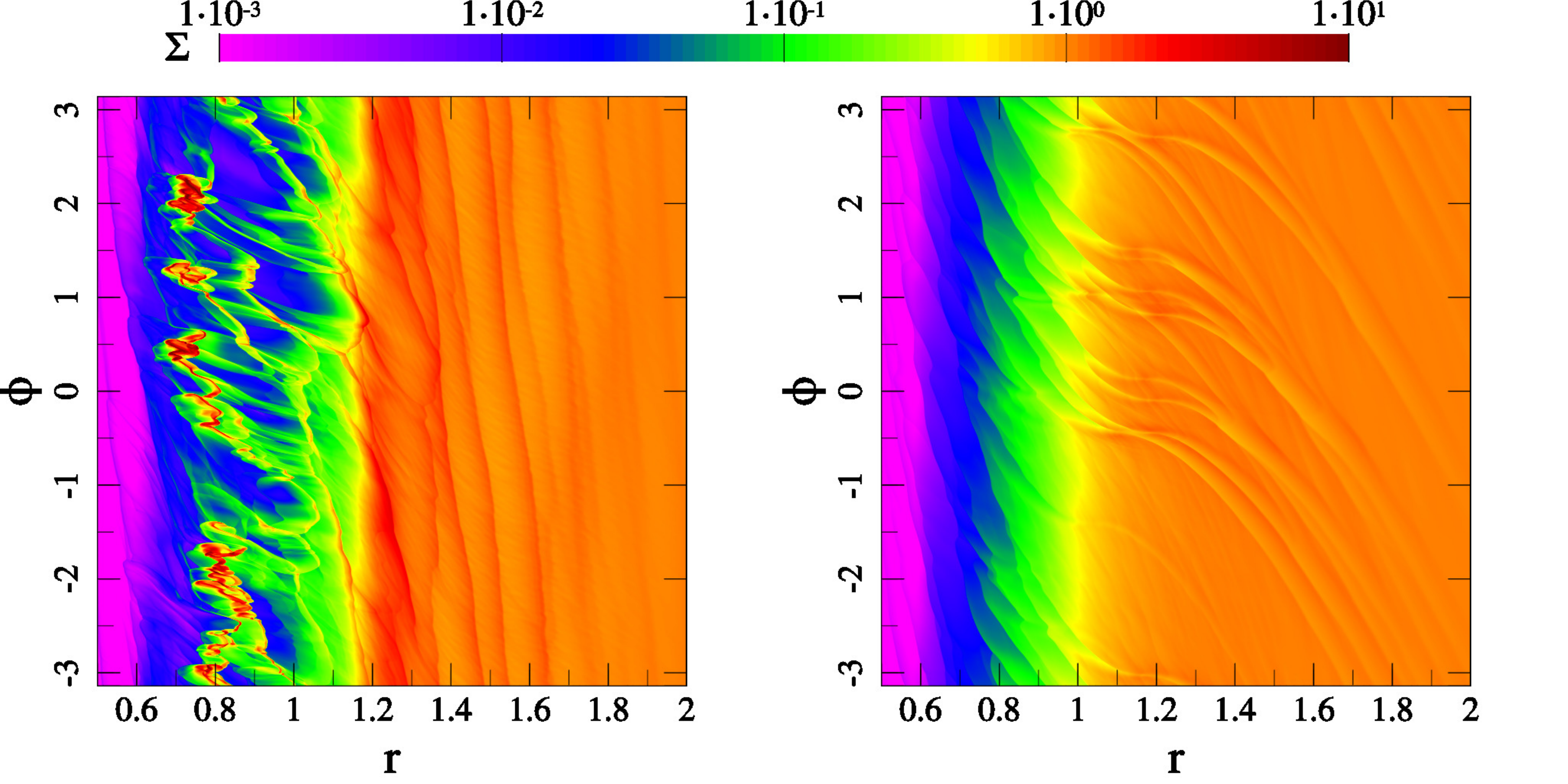}
\caption{Snapshots of our simulations for $(\Bo, \cs)=(0.2,0.02)$ on the left and $(\Bo, \cs)=(0.1,0.05)$ on the right, taken at $t=100$ orbits. Surface density is shown in logarithmic scale. The simulation on the left, belonging to region I of \figref{fig:para}, shows very high local surface density, an effect we describe as ''clumping''. On the right, belonging to region II of \figref{fig:para}, shows 6 vortices with different orbital frequencies but all lining up near $r=1.1\sim1.2$. Each of these vortices launches two pairs of spiral arms.}
\label{fig:sim}
\end{figure*}

\begin{figure*}[]
\includegraphics[width=1.99\columnwidth]{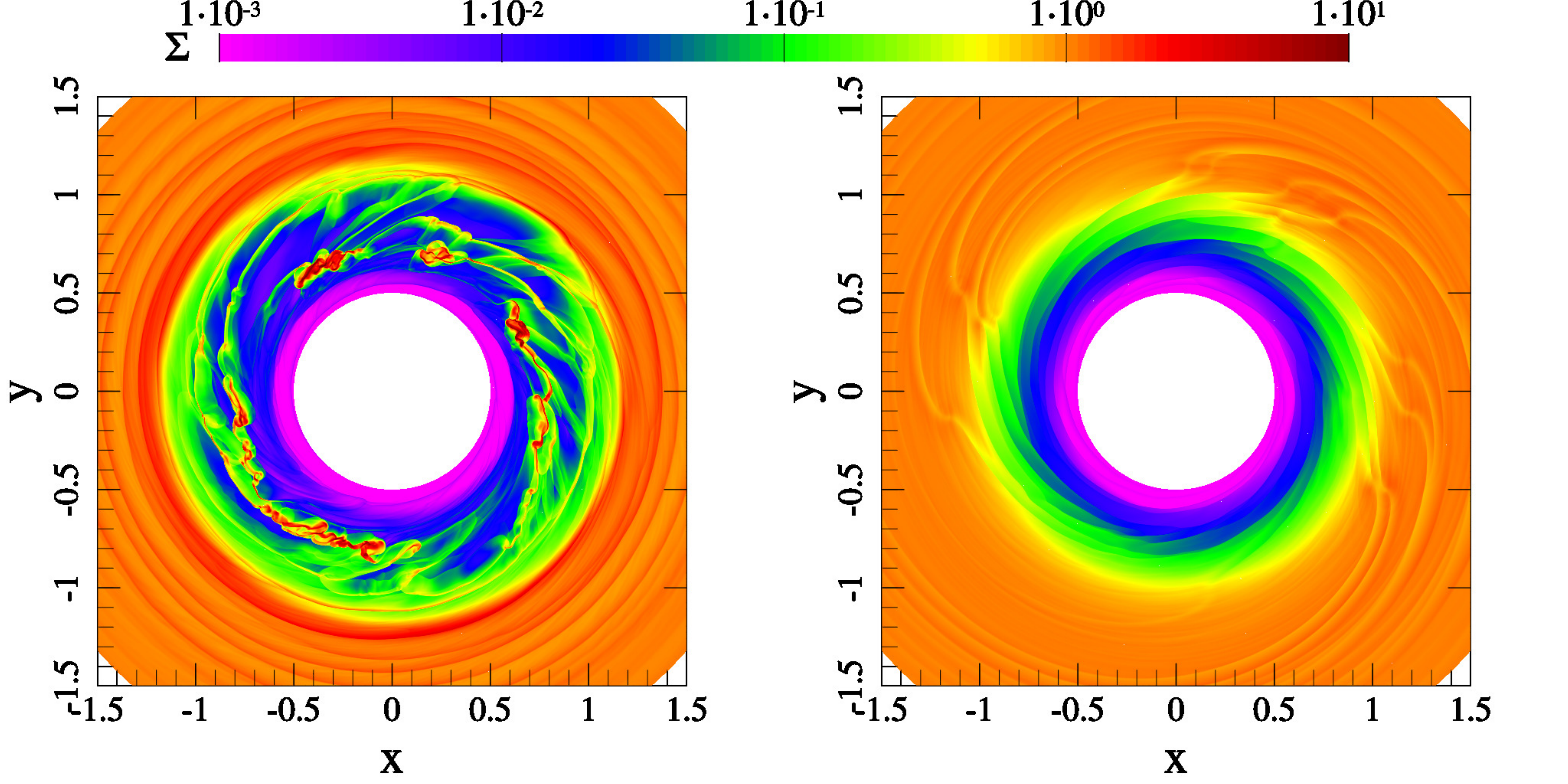}
\caption{Cartesian view of \figref{fig:sim}.}
\label{fig:sim_xy}
\end{figure*}

Nonlinear evolution is what separates region I from region II \footnote{Region III is omitted from the discussion to maintain focus on IRI. We refer the reader to the literature, e.g. \citet{RWI3,Meheut2012,Lin2013}, for detailed studies on the nonlinear evolution of RWI.} in \figref{fig:para}. Figures \ref{fig:sim} and \ref{fig:sim_xy} shows the simulation snapshots for the same two sets of parameters in the left and middle panel of \figref{fig:sim_mod} and \ref{fig:ana_mod}. The left panel, which belongs to region I with $(\Bo,\cs)=(0.2,0.02)$, shows local regions of very high surface density, exceeding 10 times $\Sigma_{d}$, the initial surface density of the disk defined in \eqnref{eq:sden}. This type of ''clumping'', which we define as a detection of $\Sigma>2\Sigma_{d}$ anywhere in the disk, is characteristic of region I. Note that clumping is not a necessary product of IRI, since region II is also driven by IRI. The transition from region II to I is rapid, in the sense that as we move toward the upper left corner of \figref{fig:para}, the highest local surface density quickly rises to a few tens of $\Sigma_{d}$.

To compare the two regions in detail, we use the right panel of \figref{fig:sim} as a typical case for region II. It shows a significant widening of the edge by a factor of $\sim3$. Vortices are formed along the edge and they create mild local enhancements in density. Their structure is complex as they typically launch two sets of spiral arms instead of one. The number of vortices is initially equal to the mode number of the fastest growing linear mode, but as they interact with each other, they can occasionally merge. It is unclear how many will remain in the long run since our simulations only last for 100 orbits at most.

In comparison to region II, the clumping in region I creates a much different, almost violent, nonlinear evolution. The clumps are very sharp features, with jumps over three orders of magnitude in density while their sizes are merely $\sim0.1r_0$. They are constantly formed and destroyed by disk shear over a dynamical timescale. The destroyed clumps form high density streams that are also visible in the left panel of \figref{fig:sim}. Another consequence of this clumping is that by concentrating a large amount of matter in a small region, radiation is able to penetrate further into the disk and push the edge of the disk to a higher orbit. In the same figure one can see that the edge of the disk is shifted to $\sim1.2r_0$. We speculate that the difference between regions I and II is due to the influence of gas pressure. Higher gas pressure results in vortex formation more similar to the purely hydrodynamical RWI, and as gas pressure becomes weaker compared to radiation, sharper features are created.

%==================================================================
\subsection{$\m{\tilde\tau}$ and the Instability Criterion Revisited}
\label{sec:tau_m}
%==================================================================

\begin{figure}[]
\includegraphics[width=0.99\columnwidth]{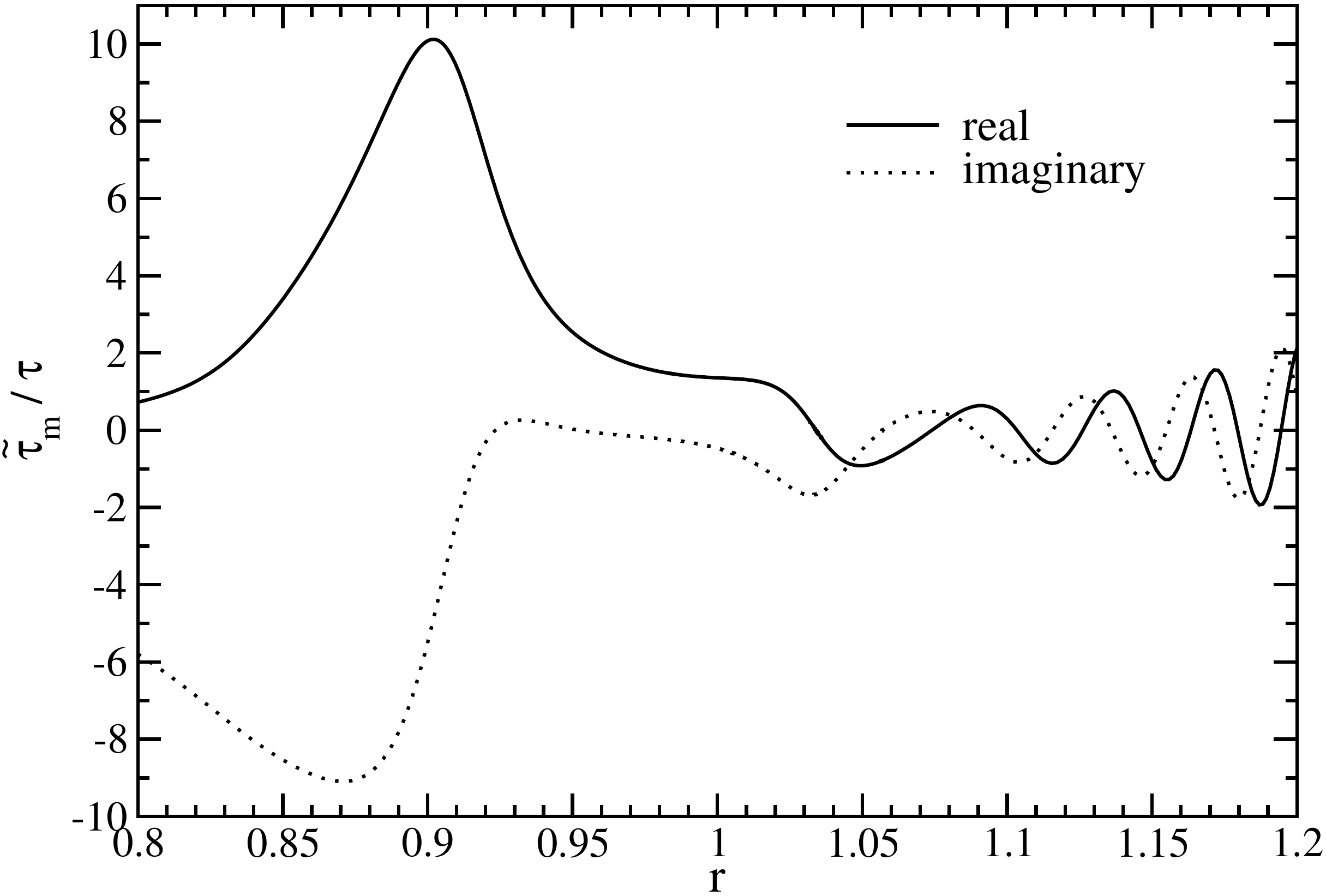}
\caption{$\m{\tilde\tau}/\tau$ for the $m=18$ mode with $(\Bo, \cs)=(0.2,0.02)$. Inside the transition region, $r\approx\{0.95,1.0\}$, the approximation $\m{\tilde\tau}\sim\tau$ is accurate to within order unity.}
\label{fig:tau_m}
\end{figure}

\begin{figure}[]
\includegraphics[width=0.99\columnwidth]{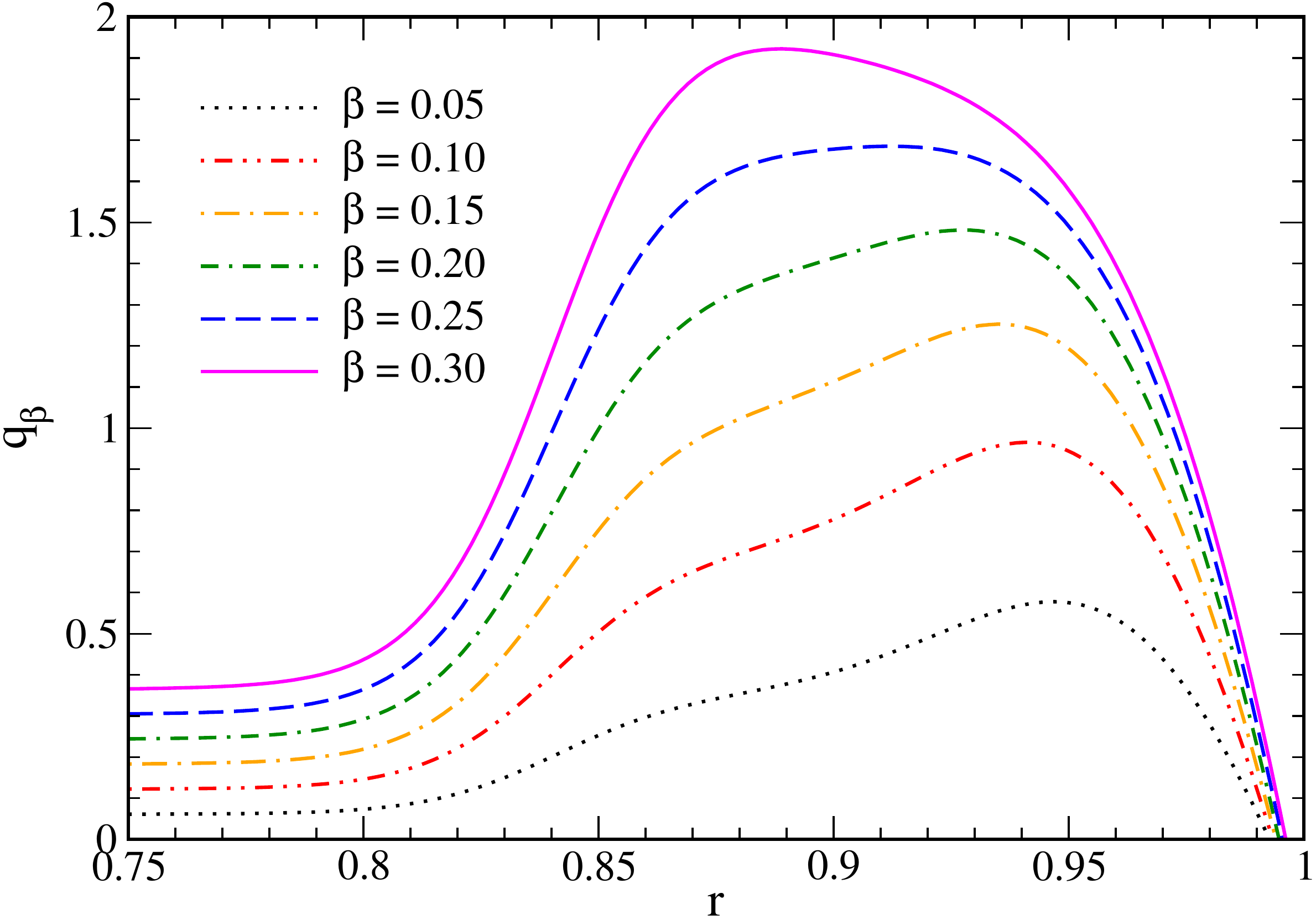}
\caption{$q_{\beta}$ from \eqnref{eq:crit3} for different values of $\Bo$. We choose $f=3$ to best match our empirical results.}
\label{fig:c_beta}
\end{figure}

In our derivation for the instability criterion in \secref{sec:crit}, we propose the crude assumption $\m{\tilde\tau}\sim\tau$. Using our semi-analytic method to obtain solutions for $\m{\eta}$, we are able to evaluate $\m{\tilde\tau}$ explicitly. For all IRI modes we have solved semi-analytically, we find $\m{\tilde\tau}/\tau>1$ within the region $r=\{r_0-\Delta r,r_0\}$, but it is never an order of magnitude above unity. For example, \figref{fig:tau_m} plots $\m{\tilde\tau}/\tau$ for the $m=18$ mode with $(\Bo, \cs)=(0.2,0.02)$. It shows that within $0.93\leq r\leq1.02$, The real part of $\m{\tilde\tau}/\tau$ is within 1 to 6, while the imaginary part is close to vanishing. Using this empirical result we can rewrite \eqnref{eq:crit2} as:
\begin{equation}
\label{eq:crit3}
q_{\beta} \approx \Bo e^{-\tau} \left( \ddlnr{\ln{\left[r\Bo\right]}} + f\tau \ddlnr{\ln{\left[r\mathscr{R}\right]}} \right) ~,
\end{equation}
where we substitute $\m{\tilde\tau}$ for $f\tau$, and $f>1$ is a number of the order of unity. Choosing $f=3$, \figref{fig:c_beta} plots $q_{\beta}$ for a few different $\Bo$. This choice of $f$ puts the threshold for instability ($q_{\beta}>1$) at around $\Bo=0.1$, similar to our empirical results.

$q_\beta$ becomes negative as soon as $\tau>1$ ($r>r_0=1$ for our disk model), because the exponential factor in $\mathscr{R}$ quickly drives its gradient negative. This implies that the instability must originate from the $\tau<1$ region. Along the same line, we find the corotation radii of IRI modes to be within $r_0$, as shown in \tabref{tab:ana}.

%==================================================================
\section{Conclusions and Discussions}
\label{sec:disc}
%==================================================================

We demonstrated that IRI can operate at an inner disk edge where there is a transition from being radially transparent to opaque. A local criterion for axisymmetric instability was derived (\eqnref{eq:crit2}). For our given disk model we computed the linear modal growth rates for $\Bo$ varying from $0$ to $0.3$, and $\cs$ from $0.02$ to $0.06$. We found growth rates ranging from $10^{-2}$ to $10^0~t_{\rm dyn}^{-1}$ (\figref{fig:para}). The fastest rates were found for the largest $\Bo$ and smallest $\cs$. We empirically determined that the threshold for IRI is $\Bo\sim0.1$ when $\Delta r=0.05$, with a weak dependence on $\cs$. For a wider edge, $\Delta r=0.1$, this threshold rises to $\Bo\sim0.25$. We note that this implies the threshold can be lowered by reducing $\Delta r$; however, at the same time $\cs$ must also be lowered for IRI to dominate over other forms of instability that may be triggered by the sharpness of the edge, such as RWI and Rayleigh instability. We employed two independent approaches to obtain the growth rates of the linear modes: simulating the disks numerically using \texttt{PEnGUIn}, and solving the linearized equations semi-analytically. Their excellent agreement lends confidence in our results. Moreover, we discovered a parameter space, labeled region I in \figref{fig:para}, where ''clumping'' occurs. There one can find over 10 times the local surface density enhancements in the nonlinear evolution of IRI.

\subsection{Connection to Physical Disks}

Our disk model is inspired by transitional disks \citep[e.g.][]{Calvet05,Espaillat07,OB1a, ResolvedImages}. The inner edges of these disks are currently unresolved by observation, but theoretical work has shown that the sharpness of disk edges created by X-ray photoevaporation \citep[e.g.][]{Owen2010} is similar to that described by our \eqnref{eq:sden} with $\Delta r=0.05$ (compare our \figref{fig:sden} to Figure 2 of \cite{Owen2013}). If a transitional disk undergoes IRI, the asymmetric structure at the inner edge will create an azimuthal variation in shadowing. \citet{Warps} showed that this can lead to a significant variation in disk emission. Indeed, some variability in the infrared emission of transitional disks has been reported by \citet{LRLL31Vari1}, \citet{LRLL31Vari2}, and \citet{SpitzerDisks}. 

On the other hand, IRI is by no means limited to circumstellar disks. AGN accretion disks, for example, can be subjected to IRI if there are any sharp jumps in density and/or opacity, such as the inner edges of the board-line regions. IRI can potentially generate the stochastic asymmetry, which is used to explain the variability in the double-peaked Balmer emission lines in radio-loud AGNs \citep{AGN08}. We note that the dynamics in AGN accretion disks are considerably more complicated since they do not have a point-like light source.

\subsection{Implications of ''Clumping''}

The ''clumping'' found in a part of our parameter space (\figref{fig:para}) opens new possibilities for IRI. For instance, very high density regions in protoplanetary disks may be favorable environments for the formation of planetary cores. The density of individual clumps may even become high enough to trigger gravitational instability at the inner edges of massive disks. One should be cautious to interpret the enhancement factors reported as realistic, however, since it is only one disk model that we have studied.

The clumping also leads to a possibility of preventing inward dust migration. \citet{DD11} demonstrated that while radiation pressure can initially push dust outward and form a dust wall, the wall eventually succumbs to the global accretion flow and migrates inward. If this wall becomes unstable due to IRI, clumping can occur, effectively creating ''leakage'' within the wall, allowing radiation to push dust further back. The true behavior of these dust walls is important to understand disks where inner clearings have been observed, such as transitional disks. Dynamical interactions between radiation, dust, and gas must be considered for this kind of study.

\subsection{Outlook}

There are three main aspects of our model that we feel would benefit greatly from a more realistic treatment. First, our model ignores the vertical dimension. A notable difference from 2D to 3D is that the location of the inner edge of a disk, defined as the $\tau=1$ point, would become a function of height, spreading over a distance of $\sim h$. One possible consequence is that IRI would generate a vertical circulation at the inner edge, which would dilute the opacity in the midplane and allow radiation pressure to penetrate further into the disk. Additionally, in a flared disk, radiation pressure is exerted on the photosphere of the entire disk rather than just the inner edge. On the other hand, because of dust settling, we expect the value of $\Bo$ in the photosphere to be smaller than the midplane, making it even more difficult to reach the $\Bo\approx0.1$ threshold. Nonetheless, for disks around exceptionally luminous stars, IRI can potentially operate at all radii.

Second, we assume a perfect coupling between gas and dust. In a more realistic approach, dust should be allowed to migrate with respect to gas. One expects dust to gather near the initial $\tau=1$ point, because where it is optically thin, dust migrates outward due to the effect of radiation pressure, and in the optically thick disk, dust migrates inward due to gas drag. This behavior of dust is described in Section 3 of \citet{TA2001}. The buildup of a dust wall is almost certain to trigger IRI due to its large $\Bo$ gradient.

Lastly, we lack a realistic treatment for radiative transfer. As the disk crosses from being radially transparent to opaque, the midplane of the disk also transitions from being heated directly by irradiation, to passively by the irradiated atmosphere. Consequently the midplane temperature should be decreasing across the disk edge. This is not captured by our globally isothermal assumption. Additionally, the clumps we find in some of our nonlinear results are sufficiently dense that they are optically thick. With our isothermal treatment, they remain the same temperature as their surroundings, while in truth these clumps should be capable of shielding themselves from irradiation and creating a non-trivial internal temperature structure. Whether this is an effect that aids or inhibits their formation and survival requires future investigation.

%==================================================================
\section*{Acknowledgments}
%==================================================================

We thank Yanqin Wu, Chris Matzner, and Eugene Chiang for helpful feedback that substantially improved this manuscript. We also thank James Owen for insightful discussions. JF owes his gratitude to the Queen Elizabeth II Graduate Scholarship in Science and Technology. We gratefully acknowledge support from the Discovery Grant by the Natural Sciences and Engineering Research Council of Canada.

\appendix
\section{Numerical Method for Solving the Linearized Equations}
\label{append}
%==================================================================
In this Appendix, we document our method for solving \eqnref{eq:2ndeq} numerically. To begin, note that \eqnref{eq:2ndeq} can only be numerically integrated in the direction of increasing $r$ because of the integral in the fourth term. In principle, it is possible to simply do this integration and find the value of $\omega$ that best matches the desired outer boundary condition. This is impractical, however, because any slight error in $\omega$ leads to a diverging behavior of $\m{\eta}$ at the outer boundary. A better method is to integrate \eqnref{eq:2ndeq} simultaneously from the inner boundary outward, and the outer boundary inward, and find the $\omega$ that results in a match of the two functions at some intermediate radius $r_{\rm mid}$. To accomplish this, we first define
\begin{equation}
\label{eq:def_y}
\m{y} \equiv \int^r_0 \frac{\m{\eta}}{\cs^2}\frac{{\rm d}\tau}{\dr'}\dr' ~,
\end{equation}
and then differentiate \eqnref{eq:2ndeq} with respect to $r$:
\begin{equation}
\label{eq:3rdeq}
\frac{\partial^3\m{y}}{\partial r^3}+a'(r)\frac{\partial^2\m{y}}{\partial r^2}+b'(r)\ppr{\m{y}}+c'(r)\m{y} = 0 ~,
\end{equation}
where
\begin{align}
\nonumber
a' &\equiv a - 2\ddr{\ln{g}},\\
\nonumber
b' &\equiv b - a\ddr{\ln{g}} + 2\left(\ddr{\ln{g}}\right)^2 - \frac{1}{g}\frac{{\rm d}^2 g}{\dr^2},\\
\nonumber
c' &\equiv c g,\\
\nonumber
g &\equiv \frac{1}{\cs^2}\ddr{\tau}.
\end{align}

Thus we can now numerically integrate \eqnref{eq:3rdeq} in both directions, and recover $\m{\eta}$ from $\m{y}$. The boundary conditions can be approximated using the WKB method. The WKB form for $\m{y}$ is
\begin{align}
\label{eq:WKB1}
\m{y} &= R(r)e^{i\int^r_0 k\dr'} ~,\\
\label{eq:WKB2}
\ppr{\m{y}} &\simeq ik\m{y} ~,
\end{align}
where $R(r)$ is a slowly varying function and $k$ is the complex wave number that satisfies $|kr|\gg1$. Substituting \eqnref{eq:WKB1} and \eqnref{eq:WKB2} into \eqnref{eq:3rdeq} we get the following algebraic equation for $k$:
\begin{equation}
\label{eq:3rdkeq}
k^3 - i a' k^2 - b' k + ic' = 0 ~.
\end{equation}
The three solutions of \eqnref{eq:3rdkeq} correspond to the inward traveling (${\rm Re}(k)<0$), outward traveling (${\rm Re}(k)>0$), and a third solution that does not exist in the conventional WKB approximation. In fact, it has $|kr|\ll1$, effectively rendering $\m{y}$ a constant, which violates the approximation of a tightly winding wave. To accommodate for this solution, we generalize \eqnref{eq:WKB1} to allow for a constant offset:
\begin{equation}
\label{eq:WKB3}
\m{y} = R(r)e^{i\int^r_0 k\dr'} + C ~.
\end{equation}
Substituting this into \eqnref{eq:3rdeq}, we obtain:
\begin{equation}
\label{eq:3rdk_mod}
k^3 - i a' k^2 - b' k + ic'\left(\frac{\m{y}}{\m{y}-C}\right) = 0 ~.
\end{equation}
In the optically thin and thick limits, $c'$ becomes arbitrarily small, and since the $|kr|\ll1$ solution is already incorporated into the constant offset $C$, the last term can be dropped, giving back the usual quadratic form:
\begin{equation}
\label{eq:2ndkeq}
k^2 - i a' k - b' = 0,
\end{equation}
which gives the expected incoming and outgoing solutions for tightly winding waves. We apply the radiative boundary condition, assuming no wave is entering the domain from the boundaries. The other unknowns remaining in \eqnref{eq:WKB3} are $R$ and $C$. For clarity, we will denote variables associated with the solution integrated from the inner boundary with the subscript ''in'', and the those from the outer boundary with ''out''. 

Recall that $\m{y}$ is in fact the integral of the perturbation (\eqnref{eq:def_y}). At the inner boundary, this quantity is small since inward of the boundary there is only a traveling wave, so we set $C_{\rm in} = 0$. We choose $R_{\rm in} = 1$, while $R_{\rm out}$ and $C_{\rm out}$ are determined by the following iterative formulas:
\begin{align}
R_{\rm out}^{i+1} &= R_{\rm out}^{i} \ddrr{y_{m,\rm in}^{i}} \left(\ddrr{y_{m,\rm out}^{i}}\right)^{-1},\\
C_{\rm out}^{i+1} &= C_{\rm out}^{i} + y_{m,\rm in}^{i} - y_{m,\rm out}^{i},
\end{align}
where $i$ is the current iterative step, the $\m{y}$ and its derivatives are evaluated at $r_{\rm mid}$. Convergence typically requires tens or even hundreds of iterations, which is the primary reason for the large amount of computational time required for this method. Lastly, we find the eigenvalue $\omega$ by minimizing the following function, evaluated at $r_{\rm mid}$:
\begin{equation}
f = \left(\frac{{\rm Re}\left(\ddr{y_{m,\rm out}}\right)-{\rm Re}\left(\ddr{y_{m,\rm in}}\right)}{{\rm max}\left[\left|{\rm Re}\left(\ddr{y_{m,\rm out}}\right)\right|,~\left|{\rm Re}\left(\ddr{y_{m,\rm in}}\right)\right|\right]}\right)^2 + \left(\frac{{\rm Im}\left(\ddr{y_{m,\rm out}}\right)-{\rm Im}\left(\ddr{y_{m,\rm in}}\right)}{{\rm max}\left[\left|{\rm Im}\left(\ddr{y_{m,\rm out}}\right)\right|,~\left|{\rm Im}\left(\ddr{y_{m,\rm in}}\right)\right|\right]}\right)^2 ~.
\end{equation}

We use an eighth-order Runge-Kutta method with adaptive step-size control for the numerical integration. We set $r_{\rm mid}=1$, the inner boundary at $r_{\rm in}=0.3$, and the outer at $r_{\rm out}=4$. Minimizing $f$ is also very time consuming because we employ a random sampling method: first we bracket the minimum within a range of likely values for the real and imaginary part of $\omega$, then we randomly select $\omega$ within the chosen range, and narrow down the field by preferentially choosing values closer to where $f$ is below a certain threshold. This time consuming method is ultimately superior to methods that involve descending along the gradient of $f$, because of the numerous local minima that exist.
%==================================================================
\bibliographystyle{apj}
\bibliography{Lit}
%==================================================================
\end{CJK*}
\end{document}